\documentclass[11pt]{article}
\usepackage{amssymb,slashed,graphicx,color,epsf,cite}
\usepackage[vcentermath]{youngtab}
\usepackage{blkarray}
\usepackage{amsmath}
\usepackage{comment}
\usepackage{lscape}
\def\l{L}

\def\tr{{\rm tr}}

\def\nn{\nonumber}

\def\be{\begin{equation}}
\def\ee{\end{equation}}
\def\ben{\begin{displaymath}}
\def\een{\end{displaymath}}
\def\bea{\begin{eqnarray}}
\def\eea{\end{eqnarray}}

\makeatletter \@addtoreset{equation}{section} \makeatother

\def\cJ{{\cal J}}
\def\cL{{\cal L}}

\def\cO{{\cal O}}
\def\cP{{\cal P}}

\newcommand{\w}[1]{\\[0.#1cm]}

\def\eq#1{(\ref{#1})}

\def\tr{{\rm tr} }

\def\tQ{{\widetilde Q}}
\def\wt{\widetilde}

\begin{document}

\begin{flushright}
\hfill{MI-TH-1941}\\

\end{flushright}

\vspace{25pt}

\begin{center}

{\bf \Large {A New Class of Ghost and Tachyon Free}}\\
 \vspace{5pt} {\bf \Large {Metric Affine Gravities }}

\vspace{0.3in}

{\large R. Percacci$^1$ and E. Sezgin$^2$}

\vspace{0.3in}

\small{\textit{$^1${\it  SISSA, via Bonomea 265, 34136 Trieste, Italy
and INFN, Sezione di Trieste}}}

\vspace{10pt}

\small{$^2${\it George and Cynthia Woods Mitchell  Institute
for Fundamental Physics and Astronomy,\\
Texas A\&M University, College Station, TX 77843, USA}}

\vspace{40pt}

\end{center}

\vskip 0.5in

\baselineskip 16pt

\begin{abstract}{ We construct the spin-projection operators for a theory containing
a symmetric two-index tensor and a general three-index tensor.
We then use them to analyse, at linearized level, 
the most general action for a metric-affine theory of gravity
with terms up to second order in curvature, which depends on 28 parameters.
In the metric case we recover known results.
In the torsion-free case, we are able to determine the most general
class of theories that are projective invariant,
contain only one massless spin 2 and no spin 3,
and are free of ghosts and tachyons.}
\end{abstract}

\vspace{15pt}

\thispagestyle{empty}

\pagebreak


\tableofcontents

\newpage

\section{Introduction}

Metric-Affine Gravity (henceforth MAG)
is a broad class of theories of gravity based on independent metric (or tetrad) and connection.
The study of MAG has a long history
\cite{Hehl:1994ue,Blagojevic:2013xpa}.
A general linear connection will have torsion and non-metricity.
In the literature, more attention has been given
to theories with torsion, but recently there has been a great deal of interest
for MAGs with non-metricity, see e.g.
\cite{Jarv:2018bgs,Aoki:2019rvi,Shimada:2018lnm,BeltranJimenez:2019tjy,BeltranJimenez:2019acz,Latorre:2017uve,Delhom:2019wir,Conroy:2017yln,Iosifidis:2018zwo,Iosifidis:2018zjj,Alvarez:2018nxc}.

There can be many reasons to study such theories.
The main reason for our interest in MAG is its relation to quadratic gravity
\footnote{By quadratic gravity we mean theories with action containing terms linear
and quadratic in the Riemann tensor.}
and its similarity to gauge theories of the fundamental interactions.
Quadratic gravity is known to be renormalizable \cite{stelle} 
and asymptotically free \cite{avrabar} but {\it prima facie} not unitary,
as expected of a theory with a kinetic term with four derivatives.
There have been many proposals to circumvent this problem,
but none has proven entirely convincing
\cite{salam,Tomboulis,floper4,Mannheim:2006rd}.
More recent progress has been reported in
\cite{Anselmi:2018ibi,Anselmi:2018tmf,Donoghue:2019fcb}.
In spite of this, there has been a revival of interest in quadratic gravity,
especially in connection with the possibility of realizing scale invariance
at high energy
\cite{Einhorn:2014gfa,Salvio:2014soa,holdom,AKKLR,donhdg}.

MAG is closely related to quadratic gravity, since it can be rewritten as
quadratic gravity coupled to a specific matter type.
Let $A$ denote a general linear connection and $F$ its curvature;
also, let $\Gamma$ be the Levi-Civita connection and $R$ its curvature.
Splitting $A=\Gamma+\phi$, where $\phi$ is a general
three-index tensor, an action of the form $\int(F+F^2)$ becomes,
schematically
\be
\int\left[R+\phi^2+(R+\nabla\phi+\phi^2)^2 \right]\ .
\label{Rphi}
\ee
In this way one can study large classes of theories of gravity and matter
with special geometrical features.
\footnote{As an example let us mention here Weyl geometry, where $\phi$ is constructed in terms of a vector field.
This theory has been revisited recently in \cite{Ghilencea:2018dqd}.}
In MAG the kinetic terms contain only two derivatives, but ghosts
are still generically present, due to the indefiniteness of the
quadratic form $F^2$.
Thus, much of the discussion that is going on
for quadratic gravity could be applied also to MAG.
However, the status of MAG is much less understood.

It is thus of obvious interest to determine what special classes of MAGs
could be free of ghosts and tachyons.
In the metric case, the most general ghost and tachyon-free theories
not containing accidental symmetries
\footnote{By accidental symmetry we mean a gauge symmetry that
is present in the linearized action but not in the full action}
have been determined in \cite{Sezgin:1979zf,sezgin2}.
It was based on the use of spin projectors for a general two-index tensor
and a three-index tensor, antisymmetric in one pair.
\footnote{This is due to the use of the vierbein formalism. The general two-index tensor is the linearized vierbein
and the three-index tensor is the linearized spin connection.}
This has been extended to to include parity-violating terms
\cite{Karananas:2014pxa,Blagojevic:2018dpz}
and a more detailed analysis of a large number of cases
including also accidental symmetries has been given recently in
\cite{Lin:2018awc}.
A broader analysis of the spectrum of a Poincar\'e gauge theory 
has been given in \cite{baikov}, where a class of ghost- and tachyon-free models were obtained.
The purpose of this paper is to give the tools that are necessary
to address this problem for general MAG, 
containing both torsion and non-metricity,
and to exhibit a new class of ghost- and tachyon-free theories
with non-metricity. 

The relation of MAG to gauge theories of fundamental interactions
is best understood if one uses arbitrary frames in the tangent bundle.
The theory is then seen to have a local gauge invariance under 
diffeomorphisms and under local $GL(4)$ transformations, 
but it is in a Higgs phase 
\cite{percacci2,Kirsch:2005st,Leclerc:2005qc,cuiaba}.
The frame field, the metric and the connection are all independent,
with the first two playing the role of Goldstone bosons.
The gauge $GL(4)$ is ``spontaneously broken'' to the trivial group
and the connection (or more precisely the difference between the connection
and the Levi-Civita connection) becomes massive.

This formalism is not well-suited for practical applications
because it contains a large number of redundant fields 
(essentially, the 16 components of the frame field).
In a linearized analysis one would discover that these fields
are all part of the kernel of the kinetic operator and
can be gauge-fixed to be zero.
It is convenient instead to work from the start
with a formalism that contains the smallest number of fields.
This is the standard formulation in terms of a metric $g_{\mu\nu}$
and an independent connection $A_\lambda{}^\mu{}_\nu$.
In this formalism the only gauge freedom is the diffeomorphism group
and one cannot reduce the number of fields further
while preserving locality.
\footnote{Except for the possible choice of unimodular gauge,
see \cite{Gielen:2018pvk}.}
It is important, however, to keep in mind that this is just a
gauge-fixed version of the general $GL(4)$ formulation,
and is gauge equivalent to the vierbein formulation.

In the following we start from the most general MAG action which contains 28 free parameters, and determine the conditions under
which it has additional symmetries under shifts of the connection.
We then determine the spin projection operators for
the fields that appear in the linearized action,
which facilitate the inversion of the wave operator 
to obtain the propagator for each spin sector.
We then specialize these results to the case of theories
with metric or torsion-free connections.
In the latter case we determine a six-parameter family of
theories that are ghost- and tachyon-free, propagating
a massless graviton and massive spin $2^-$ or $1^+$ 
states with distinct masses.

\section{Metric affine gravity}

\subsection{The action}

In the model we shall consider, the independent dynamical variables 
are the metric $g_{\mu\nu}$ of signature $-+\ldots+$ and a linear connection 
$A_\mu{}^\rho{}_\sigma$ 
The curvature is defined as
\begin{eqnarray}
&F_{\mu\nu}{}^\rho{}_\sigma=\,\partial_\mu A_\nu{}^\rho{}_\sigma-\partial_\nu 
A_\mu{}^\rho{}_\sigma
+A_\mu{}^\rho{}_\tau A_\nu{}^\tau{}_\sigma
-A_\nu{}^\rho{}_\tau A_\mu{}^\tau{}_\sigma\ ,
\end{eqnarray}
whereas torsion and non-metricity are defined by \footnote{Note that the torsion tensor is antisymmetric in its first and third indices. This is not to be confused with the convention used widely in the supergravity literature where it is antisymmetric in its first two indices instead.}
\bea
T_\mu{}^\alpha{}_\nu&=&A_\mu{}^\alpha{}_\nu-A_\nu{}^\alpha{}_\mu\ ,
\\
Q_{\lambda\mu\nu}&=&-\partial_\lambda g_{\mu\nu}
+A_\lambda{}^\tau{}_\mu g_{\lambda\nu}
+A_\lambda{}^\tau{}_\nu g_{\mu\lambda}
\ .
\eea
As an action we take
\bea
S(g,A) &=&-\frac12\int d^dx\ \sqrt{|g|}\,\Big[ -a_0 F+ F^{\mu\nu\rho\sigma} \big( c_1 F_{\mu\nu\rho\sigma} 
+ c_2 F_{\mu\nu\sigma\rho} 
+ c_3 F_{\rho\sigma\mu\nu} 
+ c_4 F_{\mu\rho\nu\sigma} 
\nn\w2
&&  
+ c_5 F_{\mu\sigma\nu\rho} 
+ c_6 F_{\mu\sigma\rho\nu} \big)
+ F^{(13)\mu\nu} \big(c_7 F^{(13)}_{\mu\nu} + c_8 F^{(13)}_{\nu\mu} \big)
+ F^{(14)\mu\nu} \big( c_9 F^{(14)}_{\mu\nu} 
+ c_{10} F^{(14)}_{\nu\mu}\big) 
\nn\w2
&& 
+ F^{(14)\mu\nu}\big(c_{11} F^{(13)}_{\mu\nu}
+ c_{12} F^{(13)}_{\nu\mu} \big)
+F^{\mu\nu}\big(c_{13} F_{\mu\nu}
+ c_{14} F^{(13)}_{\mu\nu}
+ c_{15} F^{(14)}_{\mu\nu}\big)
+c_{16}F^2
\nn\w2
&& 
+ T^{\mu\rho\nu} \big(a_1 T_{\mu\rho\nu} + a_2 T_{\mu\nu\rho}\big)  + a_3 T^\mu T_\mu
+Q^{\rho\mu\nu}\big( a_4 Q_{\rho\mu\nu} 
+ a_5 Q_{\nu\mu\rho}\big)  
\nn\w2
&&
+ a_6 Q^\mu Q_\mu + a_7 \tQ^\mu \tQ_\mu + a_8 Q^\mu \tQ_\mu
 + a_9 T^{\mu\rho\nu} Q_{\mu\rho\nu}  
+ T^\mu \left( a_{10} Q_\mu 
+ a_{11}\tQ_\mu \right)
\Big]\ ,
\label{action}
\eea
where
\bea
T_\mu &:=&  T_\lambda{}^\lambda{}_\mu\ ,\qquad Q_\mu := Q_{\mu\lambda}{}^\lambda\ ,\qquad \tQ_\mu := Q_\lambda{}^\lambda{}_\mu\ ,
\nn\w2
F_{\mu\nu} &:=& F_{\mu\nu\lambda}{}^\lambda\ ,\quad F_{\mu\nu}^{(14)} := F_{\lambda\mu\nu}{}^\lambda\ ,\quad F_{\mu\nu}^{(13)} := F_{\lambda\mu}{}^\lambda{}_\nu\ ,\quad
F := F_{\mu\nu}{}^{\mu\nu}\ .
\eea
Note that there are two ``pseudo-Ricci'' tensors $F_{\mu\nu}^{(13)}$
and $F_{\mu\nu}^{(14)}$,
without symmetry properties, and one ``pseudo-Ricci scalar" that we denote $F$.
The Einstein-Hilbert action is described by the $a_0 g^{\mu\nu} F^{(13)}_{\mu\nu}$ term.
The action contains $28$ parameters, namely ($a_0$, $a_1$,...,$a_{11}$, $c_1$,...,$c_{16}$). In $d=4$, however, the combination
\be
 F_{\mu\nu\rho\sigma} F^{\rho\sigma\mu\nu} 
- F^{(13)}_{\mu\nu} F^{\nu\mu(13)}- F^{(14)}_{\mu\nu}F^{(14)\nu\mu}
+ 2 F^{(13)}_{\mu\nu} F^{\nu\mu(14)} + F^2\ ,
\label{gaussbonnet}
\ee
which reduces to the Gauss-Bonnet integrand in the Riemannian case,
does not contribute at quadratic level when expanding around flat space.
Indeed, in Weyl geometry (i.e. if the non-metricity is of the form 
$Q_{\lambda\mu\nu}=v_\lambda g_{\mu\nu}$), it is a total derivative
\cite{Babourova:1996hy}.
In the presence of tracefree non-metricity, it is not a total derivative \cite{Janssen:2019uao}, but in flat space it
only gives cubic and quartic interactions .
Thus, for the purposes of our analysis, one parameter is redundant.
Turning to the action \eq{action}, it is convenient to express it as
\bea
S(g,A) &=&-\frac12\int d^dx\ \sqrt{|g|}\,\Big[ 
G^{\mu_1\dots \mu_4, \nu_1\dots \nu_4}  F_{\mu_1\dots \mu_4} F_{\nu_1\dots \nu_4}
 \nn\\
&&
+ A^{\mu_1\mu_2\mu_3, \nu_1\nu_2\nu_3}  T_{\mu_1\mu_2\mu_3} T_{\nu_1\nu_2\nu_3}
 + B^{\mu_1\mu_2\mu_3, \nu_1\nu_2\nu_3}  Q_{\mu_1\mu_2\mu_3}Q_{\nu_1\nu_2\nu_3}
 \nn\\
&&
+  C^{\mu_1\mu_2\mu_3, \nu_1\nu_2\nu_3} T_{\mu_1\mu_2\mu_3}Q_{\nu_1\nu_2\nu_3}
\Big]\ .
\eea
The tensors $G$, $A$, $B$ and $C$ inherit the symmetries of the objects
they are contracted with.
Furthermore, $G$, $A$ and $B$ are also symmetric under
the interchange of the first half of indices with the second half.
In the following expressions, symmetrizations that are not
already manifest are indicated:
\footnote{In our conventions, the (anti) symmetrizations are always with unit strength, e.g. $X_{[a} Y_{b]}= \frac12 (X_a Y_b -X_bY_a)$.}
\bea
G_{\mu_1...\mu_4}{}^{\nu_1...\nu_4} &=& 
\Big[\delta_{\mu_1}^{\nu_1}\delta_{\mu_2}^{\nu_2} \left( c_1\, \delta_{\mu_3}^{\nu_3}\delta_{\mu_4}^{\nu_4}
+c_2\, \delta_{\mu_3}^{\nu_4}\delta_{\mu_4}^{\nu_3}\right) 
+c_3\,\delta_{\mu_1}^{\nu_3}\delta_{\mu_2}^{\nu_4}\delta_{\mu_3}^{\nu_1}\delta_{\mu_4}^{\nu_2}
+c_4\, \delta_{\mu_1}^{\nu_1}\delta_{\mu_2}^{\nu_3}\delta_{\mu_3}^{\nu_2}\delta_{\mu_4}^{\nu_4}
\nn\\
&& +\delta_{\mu_1}^{\nu_1}\delta_{\mu_2}^{\nu_4}
\left(c_5\, \delta_{\mu_3}^{\nu_2}\delta_{\mu_4}^{\nu_3}
+c_6\, \delta_{\mu_3}^{\nu_3}\delta_{\mu_4}^{\nu_2}\right) +\eta_{\mu_1\mu_3}\eta^{\nu_1\nu_3}\left( c_7\, \delta_{\mu_2}^{\nu_2}\delta_{\mu_4}^{\nu_4} +c_8\, \delta_{\mu_2}^{\nu_4}\delta_{\mu_4}^{\nu_2}\right)
\nn\\
&& +\eta_{\mu_1\mu_4}\eta^{\nu_1\nu_4}\left( c_9\, \delta_{\mu_2}^{\nu_2}\delta_{\mu_3}^{\nu_3} +c_{10}\, \delta_{\mu_2}^{\nu_3}\delta_{\mu_3}^{\nu_2}\right)+\eta_{\mu_1\mu_4}\eta^{\nu_1\nu_3}\left( c_{11}\, \delta_{\mu_2}^{\nu_2}\delta_{\mu_3}^{\nu_4} +c_{12}\, \delta_{\mu_2}^{\nu_4}\delta_{\mu_3}^{\nu_2}\right)
\nn\\
&&+ \eta_{\mu_3\mu_4}\left( c_{13}\, \eta^{\nu_3\nu_4}\delta_{\mu_1}^{\nu_1}\delta_{\mu_2}^{\nu_2} +c_{14}\,\eta^{\nu_1\nu_3} \delta_{\mu_1}^{\nu_2}\delta_{\mu_2}^{\nu_4} + c_{15}\, \eta^{\nu_1\nu_4} \delta_{\mu_1}^{\nu_2}\delta_{\mu_2}^{\nu_3} \right)
\nn\\
&& +c_{16}\,\eta_{\mu_1\mu_3}\eta_{\mu_2\mu_4}\eta^{\nu_1\nu_3}\eta^{\nu_2\nu_4}\Big]_{[\mu_1\mu_2][\nu_1\nu_2]}\ ,
\w4
A_{\mu_1\mu_2\mu_3}{}^{\nu_1\nu_2\nu_3} &=& \Big[\delta_{\mu_1}^{\nu_1} \left(
a_1\, \delta_{\mu_2}^{\nu_2}\delta_{\mu_3}^{\nu_3} 
+a_2\, \delta_{\mu_2}^{\nu_3}\delta_{\mu_3}^{\nu_2}\right) 
+a_3\, \eta_{\mu_1\mu_2}\eta^{\nu_1\nu_2} \delta_{\mu_3}^{\nu_3}\Big]_{[\mu_1\mu_2][\nu_1\nu_2]}\ ,
\eea
\bea
B_{\mu_1\mu_2\mu_3}{}^{\nu_1\nu_2\nu_3} &=& \Big[a_4\, \delta_{\mu_1}^{\nu_1} \delta_{\mu_2}^{\nu_2}\delta_{\mu_3}^{\nu_3} 
+a_5\, \delta_{\mu_1}^{\nu_3} \delta_{\mu_2}^{\nu_2}\delta_{\mu_3}^{\nu_1} 
+a_6\, \eta_{\mu_2\mu_3}\eta^{\nu_2\nu_3}\delta_{\mu_1}^{\nu_1}
\nn\\
&& +\eta_{\mu_1\mu_2}\left(a_7\,\eta^{\nu_1\nu_2}\delta_{\mu_3}^{\nu_3} + a_8\,\eta^{\nu_2\nu_3}\delta_{\mu_3}^{\nu_1}\right)\Big]_{(\mu_2\mu_3)(\nu_2\nu_3)}\ ,
\nn\w4
C_{\mu_1\mu_2\mu_3}{}^{\nu_1\nu_2\nu_3} &=& \Big[a_9\, \delta_{\mu_1}^{\nu_1} \delta_{\mu_2}^{\nu_2}\delta_{\mu_3}^{\nu_3} +\eta^{\nu_1\nu_2} \left( a_{10}\,\eta_{\mu_2\mu_3}\delta_{\mu_1}^{\nu_3} + a_{11} \eta_{\mu_1\mu_2}\delta_{\mu_3}^{\nu_3}\right)\Big]_{(\mu_2\mu_3)[\nu_1\nu_2]}\ ,
\label{GH}
\eea
where it is understood that $G$ is to be symmetrized with respect to interchange of indices $(\mu_1...\mu_4)$ and $(\nu_1...\nu_4)$, and that $A$, $B$ and $C$
with respect to the interchange of indices $(\mu_1...\mu_3)$ and $(\nu_1...\nu_3)$.
%

\subsection{Gauge symmetries}

In general the action is invariant under the 
action of diffeomorphisms,
\bea
g'_{\mu\nu}(x') &=& \frac{\partial x^\alpha}{\partial x'^\mu}
\frac{\partial x^\beta}{\partial x'^\nu}
g_{\alpha\beta}(x)\ ,
\\
A'_\mu{}^\alpha{}_\beta(x')&=&
\frac{\partial x^\nu}{\partial x'^\mu}
\frac{\partial x'^\alpha}{\partial x^\gamma}
\frac{\partial x^\delta}{\partial x'^\beta}A_\nu{}^\gamma{}_\delta(x) 
+ \frac{\partial x'^\alpha}{\partial x^\gamma}
\frac{\partial^2 x^\gamma}{\partial x'^\mu\partial x'^\beta}\ .
\label{diffeo}
\eea
For an infinitesimal transformation $x^{\prime\mu}=x^\mu-\xi^\mu(x)$
the transformation is given by the Lie derivatives, plus
an inhomogeneous term for the connection:
\be
\delta g_{\mu\nu} = \cL_\xi g_{\mu\nu}\ ,\qquad 
\delta A_\rho{}^\mu{}_\nu =\cL_\xi A_\rho{}^\mu{}_\nu
+\partial_\rho \partial_\nu \xi^\mu\ ,
\label{lindiff}
\ee
where $\cL_\xi A_\rho{}^\mu{}_\nu
=\xi^\lambda\partial_\lambda A_\rho{}^\mu{}_\nu
+A_\lambda{}^\mu{}_\nu \partial_\rho\xi^\lambda
-A_\rho{}^\lambda{}_\nu \partial_\lambda\xi^\mu
+A_\rho{}^\mu{}_\lambda \partial_\rho\xi^\lambda$.
In four dimensions, if all the coefficients $a_i$ are zero, the action is additionally invariant under the following realization of Weyl transformations:
\be
\delta g_{\mu\nu}=2\omega g_{\mu\nu}
;\qquad
\delta A_\mu{}^\rho{}_\nu =0 \ .
\ee
This is the usual way in which Weyl transformations are realized on Yang-Mills fields, while the Levi-Civita connection transforms as:
\be
\delta \Gamma_\mu{}^\rho{}_\nu=
\partial_\mu\omega \delta^\rho_\nu
+\partial_\nu\omega \delta^\rho_\mu
-g^{\rho\tau}\partial_\tau\omega g_{\mu\nu} 
\ .
\ee

In the following we shall be interested in cases where the action is invariant under additional transformations of the connection
(see also \cite{Iosifidis:2019fsh}). The following three classes of transformations will be relevant.
First we consider the projective transformations
\be
\delta_1 A_\mu{}^\rho{}_\nu = \lambda_\mu \delta^\rho_\nu\ ,\qquad \delta_1 g_{\mu\nu}=0\ ,
\label{pt}
\ee
where $\lambda_\mu(x)$ is an arbitrary gauge parameter. Under this transformation 
\bea
\delta_1 F_{\mu\nu\rho\sigma} &=& \left( 2\nabla_{[\mu}\lambda_{\nu]} +T_\mu{}^\tau{}_\nu \lambda_\tau\right) g_{\rho\sigma} = 2\partial_{[\mu}\lambda_{\nu]} g_{\rho\sigma}\ ,
\nn\\
\delta_1 T_\mu{}^\rho{}_\nu &=& 2\lambda_{[\mu} \delta_{\nu]}^\rho\ ,\qquad \delta_1 Q_{\rho\mu\nu}=2\lambda_\rho\, g_{\mu\nu}\ .
\eea
In particular $\delta_1 F=0$. Assuming that neither torsion nor the non-metricity vanish, one finds that the action is invariant provided that
\bea
&& 2c_1+2c_2+2dc_{13}-c_{14}-c_{15}=0\ ,
\nn\\
&& c_5+2c_6+2c_7-2c_8+c_{11}-c_{12}-d c_{14}=0\ ,
\nn\\
&& 2c_4+c_5+2c_9-2c_{10}+c_{11}-c_{12}-dc_{15}=0\ ,
\nn\\
&& 2a_1+a_2+(d-1)a_3+a_9-da_{10}-a_{11}=0\ ,
\nn\\
&& 4a_4+4d a_6+2a_8+a_9-(d-1) a_{10}=0\ ,
\nn\\
&& 4a_5+4a_7+2da_8-a_9-(d-1)a_{11}=0\ .
\eea

There is a similar transformation with the second index singled out
\be
\delta_2 A_\mu{}^\rho{}_\nu = \lambda^\rho g_{\mu\nu}\ ,\qquad 
\delta_2 g_{\mu\nu}=0\ ,
\label{pt2}
\ee
under which
\bea
\delta_2 F_{\mu\nu\rho\sigma} &=& 2g_{\sigma [\nu} \nabla_{\mu]}\lambda_\rho +2g_{\sigma[\nu} Q_{\mu]\rho\tau} \lambda^\tau +2\left( Q_{[\nu\mu]\sigma} + T_{\mu\sigma\nu}\right)\lambda_\rho\ ,
\nn\w2
\delta_2 T_\mu{}^\rho{}_\nu &=& 0\ ,\quad \delta_2 Q_{\rho\mu\nu} = 2g_{\rho(\mu} \lambda_{\nu)}\ .
\label{s2}
\eea
In this case, the variation of the general action gives rise to a large number of independent structures. Then the invariance of the action requires that 
\bea
&& c_1=c_2=\ldots =c_{16}=0\ ,
\nn\\
&&  (2-d)a_0+ a_9+2a_{10}+(d+1)a_{11}  =0\ ,
\nn\\
&& (3-d)a_0+ 4a_5+8a_6+2(d+1)a_8 =0\ ,
\nn\\
&& -a_0+ 4a_4+2a_5+2(d+1)a_7+2a_8 =0\ .
\eea

Finally there is the transformation that singles out the third index
\be
\delta_3 A_\mu{}^\rho{}_\nu =  \delta^\rho_\mu\lambda_\nu\ ,\qquad \delta g_{\mu\nu}=0\ ,
\label{pt3}
\ee
under which
\be
\delta_3 F_{\mu\nu\rho\sigma} = 2 g_{\rho[\nu} \nabla_{\mu]} \lambda_\sigma +T_{\mu\rho\nu}\lambda_\sigma\ ,\quad 
\delta_3 T_\mu{}^\rho{}_\nu =
2\delta^\rho_{[\mu}\lambda_{\nu]}\ ,\quad \delta_3 Q_{\rho\mu\nu} =  2g_{\rho(\mu} \lambda_{\nu)}\ .
\ee
Once again, the variation of the general action gives rise to a large number of independent structures. Assuming that the torsion and non-metricity do not vanish, the action is invariant provided that 
\bea
&& c_1=c_2=\ldots =c_{16}=0\ ,
\nn\\
&& (d-2)a_0+ 4a_1+2a_2+2(d-1)a_3+a_9+2a_{10}+(d+1)a_{11} =0\ ,
\nn\\
&& (d-1)a_0+4a_5+8a_6+2(d+1)a_8-2a_9+2(d-1)a_{10}=0\ ,
\nn\\
&& (1-d)a_0+ 4a_4+2 a_5+2(d+1)a_7+2a_8+a_9 +(d-1) a_{11}  =0 \ .
\eea

\section{Linearization and spin projectors}

\subsection{Linearized action}

The equations of motion that come from the action (\ref{action})
have as a solution the Minkowski space
\be
g_{\mu\nu}=\eta_{\mu\nu}\ ,\quad
A_\rho{}^\mu{}_\nu=0\ .
\ee
Expanding the action around this solution,
the quadratic wave operator takes the form
\be
S^{(2)} = \frac12\int d^dq \left(\, 
A^{\lambda\mu\nu}\,{\cal O}_{\lambda\mu\nu}{}^{\tau\rho\sigma}\, A_{\tau\rho\sigma} 
+2A^{\lambda\mu\nu}\,{\cal O}_{\lambda\mu\nu}{}^{\rho\sigma}\, h_{\rho\sigma} 
+ h^{\mu\nu}\,{\cal O}_{\mu\nu}{}^{\rho\sigma}\,h_{\rho\sigma}\, \right) \ ,
\label{charlie}
\ee
where, by abuse of notation, we denote $A$ also the fluctuation, and
\bea
{\cal O}^{\mu\nu,\rho\sigma}\!\!\!\! &=&\!\!\!\!
-B^{\lambda\mu\nu,\tau \rho\sigma}q_\lambda q_\tau\ ,
\\
{\cal O}^{\lambda\mu\nu,\rho\sigma}\!\!\!\! &=&\!\!\!\!-2i
\left(A^{\lambda\mu\nu,\tau\rho\sigma}+C^{\lambda\mu\nu,\tau\rho\sigma}\right)q_\tau
+\frac{i}{2}a_0\left[\eta^{\nu\sigma}(\eta^{\lambda\rho}q^\mu-\eta^{\lambda\mu}q^\rho)
-\frac12\eta^{\rho\sigma}(\eta^{\lambda\nu}q^\mu-\eta^{\lambda\mu}q^\nu)\right]
\ ,
\nn\\
{\cal O}^{\lambda\mu\nu,\tau\rho\sigma}\!\!\!\! &=&\!\!\!\!
-4\left(G^{\kappa\lambda\mu\nu,\eta \tau\rho\sigma}q_\kappa q_\eta
+A^{\lambda\mu\nu,\tau\rho\sigma} +B^{\lambda\mu\nu,\tau\rho\sigma} + 2C^{\lambda\mu\nu,\tau\rho\sigma}\right)
+a_0\eta^{\nu\rho}\left(\eta^{\lambda\mu}\eta^{\tau\sigma}-\eta^{\lambda\sigma}\eta^{\mu \tau}\right).
\nn
\eea
This operator has a kernel consisting (at least) of the infinitesimal diffeomorphisms (\ref{lindiff}), which 
in the present case read
\be
\delta g_{\mu\nu} = \partial_\mu \xi_\nu+\partial_\nu\xi_\mu\ ,\qquad 
\delta A_{\lambda\mu\nu} =\partial_\nu \partial_\lambda \xi_\mu\ .
\label{lindiff2}
\ee
For specific values of the couplings the kernel could be larger.

\subsection{Spin projectors}
In the analysis of the spectrum of operators acting on
multi-index fields in flat space, it is very convenient to
use spin-projection operators, which can be used to
decompose the fields in their irreducible components
under the three-dimensional rotation group
\cite{rivers,barnes,aurilia}.
For a three-index tensor that is antisymmetric in one pair
of indices, the spin projectors were given in
\cite{neville,Sezgin:1979zf}.
The spin projectors for totally symmetric three-tensors
have been given also in \cite{Mendonca:2019gco}.
To the best of our knowledge, the spin projectors for a general
three-index tensor have not been given in the literature.
We thus turn to the construction of these objects.

\subsubsection{$GL(d)$-decomposition}
The space of two-index tensors can be decomposed into irreps of the group $GL(d)$, given by
symmetric and antisymmetric tensors. The projectors onto these subspaces are
\be
\Pi^{(s/a)}_{ab}{}^{ef}= \frac{1}{2}
\left(\delta_a^e\delta_b^f \pm  \delta_a^f\delta_b^e\right)\ .
\label{psa}
\ee
The finer decomposition into irreps of $SO(d-1)$ is widely used in gravity.
The corresponding treatment of three-index tensors
is algebraically more complicated.
We begin with some elementary facts about three-index tensors as representations of $GL(d)$.
In order to discuss their symmetry properties,
we will focus on the second pair of indices.
Thus when we say that $t_{cab}$ is (anti)symmetric,
without further specification, we mean $t_{cba}=\mp t_{cab}$.

The space $V$ of three-index tensors has dimension $d^3$.
The subspaces $V^{(s)}$ and $V^{(a)}$
of symmetric and antisymmetric tensors 
are invariant subspaces of dimensions
$d^2(d+1)/2$ and $d^2(d-1)/2$ respectively.
The projectors onto these subspaces are
\be
\Pi^{(s)}_{cab}{}^{def}=
\frac{1}{2}\delta_c^d
\left(\delta_a^e\delta_b^f+
\delta_a^f\delta_b^e\right)
\ , \qquad
\Pi^{(a)}_{cab}{}^{def}=
\frac{1}{2}\delta_c^d
\left(\delta_a^e\delta_b^f-\delta_a^f\delta_b^e\right)\ .
\label{sa3a}
\ee

The subspaces $V^{(ts)}$ and $V^{(ta)}$
of totally symmetric and totally antisymmetric
tensors are invariant subspaces of dimension
$d(d-1)(d-2)/6$ and $d(d+1)(d+2)/6$ respectively.
Given any tensor, one can extract its totally (anti) symmetric 
part by means of the projectors
\bea
\Pi^{(ts)}_{cab}{}^{def}&=&
\frac{1}{6}
\left(\delta_c^d\delta_a^e\delta_b^f
+\delta_c^d\delta_a^f\delta_b^e
+\delta_c^f\delta_a^d\delta_b^e
+\delta_c^f\delta_a^e\delta_b^d
+\delta_c^e\delta_a^f\delta_b^d
+\delta_c^e\delta_a^d\delta_b^f
\right)\ ,
\nn\\
\Pi^{(ta)}_{cab}{}^{def}&=&
\frac{1}{6}
\left(\delta_c^d\delta_a^e\delta_b^f
-\delta_c^d\delta_a^f\delta_b^e
+\delta_c^f\delta_a^d\delta_b^e
-\delta_c^f\delta_a^e\delta_b^d
+\delta_c^e\delta_a^f\delta_b^d
-\delta_c^e\delta_a^d\delta_b^f
\right)\ .
\label{sa3b}
\eea
The complement of $V^{(ts)}$ in $V^{(s)}$ and of $V^{(ta)}$ in $V^{(a)}$
are also invariant subspaces denoted $V^{(hs)}$ and $V^{(ha)}$
respectively. 
\footnote{``$hs$'' and ``$ha$'' stand for ``hook-symmetric'' and
``hook antisymmetric'', since these tensors have the structure
of the hook Young tableau.}
They consist of tensors that are (anti) symmetric but
have zero totally (anti) symmetric part. The projectors onto such subspaces are
\bea
\Pi^{(hs)}_{cab}{}^{def}&=&\Pi^{(s)}_{cab}{}^{def}-\Pi^{(ts)}_{cab}{}^{def} = 
\frac16 \left( 2 \delta_ c^d \delta_a^e \delta_b^f -  \delta_ b^d \delta_c^e \delta_a^f  -   \delta_ a^d \delta_b^e \delta_c^f \right)     + a\ \leftrightarrow\ b   \ ,
\nn\\
\Pi^{(ha)}_{cab}{}^{def}&=&\Pi^{(a)}_{cab}{}^{def}-\Pi^{(ta)}_{cab}{}^{def} = 
\frac16 \left( 2 \delta_ c^d \delta_a^e \delta_b^f -  \delta_ b^d \delta_c^e \delta_a^f  -   \delta_ a^d \delta_b^e \delta_c^f \right)     - a\ \leftrightarrow\ b    \ ,
\eea
Thus the decomposition of a three-index tensor in its $GL(d)$-irreducible parts is
\be
t_{cab}=t^{(ts)}_{cab}+t^{(hs)}_{cab} +t^{(ha)}_{cab}+t^{(ta)}_{cab}\ .
\ee
where
\bea
t^{(ts)}_{cab}&=&\frac{1}{6}\left(
t_{cab}+t_{cba}+t_{bca}+t_{bac}+t_{abc}+t_{acb}
\right)\ ,\nn\\
t^{(hs)}_{cab}&=&\frac{1}{6}\left(
2t_{cab}+2t_{cba}-t_{acb}-t_{abc}-t_{bca}-t_{bac}
\right)\ ,\nn\\
t^{(ha)}_{cab}&=&\frac{1}{6}\left(
2t_{cab}-2t_{cba}+t_{acb}-t_{abc}-t_{bca}+t_{bac}
\right)\ ,\nn\\
t^{(ta)}_{cab}&=&\frac{1}{6}\left(
t_{cab}-t_{cba}+t_{bca}-t_{bac}+t_{abc}-t_{acb}
\right)\ .
\eea
%

\subsubsection{$SO(d-1)$-decomposition}

A four-vector $q^a$ with $q^2\not=0$
breaks $SO(1,d-1)$ to $SO(d-1)$.
In physical applications $q^a$ has the meaning of a four-momentum.
Given $q^a$, we can decompose every other
vector in parts longitudinal and transverse to it,
by using the projectors
\be
{\hat q}^a \equiv ~q^a/\sqrt{|q^2|}\ ,\qquad 
L_a{}^b={\hat q}_a{\hat q}^b\ ,
\qquad 
T_a{}^b=\delta_a^b-\l_a^b\ .
\ee
This leads to a finer decomposition of $V$ into irreps
of the group $SO(d-1)$.
As a first step we expand the identity 
\be
\delta_c^d\delta_a^e\delta_b^f=
(T_c^d+L_c^d)(T_a^e+L_a^e)(T_b^f+L_b^f)
\ee
in eight terms. It is easy to see that the combinations
\be
TTT\ ;\quad 
TTL+TLT+LTT\ ,\quad 
TLL+LTL+LLT\ ,\quad 
LLL
\ee
(all with fixed indices) are projectors. Then consider the simultaneous eigenspaces with eigenvalue 1 of these and of the $GL(d)$
projectors introduced above. The dimensions of these spaces are given in Table \ref{t1}. The last column and the last row give the total dimension of the +1 eigenspaces of the projectors in the corresponding rows and columns. 

\bigskip
\begin{table}
\begin{center}
\begin{tabular}{|c|c|c|c|c|c|}
\hline
        & $ts$ & $hs$ & $ha$ & $ta$ & dim \\
\hline
$TTT$   & $\frac{d(d^2-1)}{6}$  & $\frac{d(d-1)(d-2)}{3}$  & $\frac{d(d-1)(d-2)}{3}$  &  $\frac{(d-3)(d-2)(d-1)}{6}$ & $(d-1)^3$ \\
\hline
$TTL+TLT+LTT$  &  $\frac{d(d-1)}{2}$ & $(d-1)^2$  & $(d-1)^2$  & $\frac{(d-2)(d-1)}{2}$  & $3(d-1)^2$\\
\hline
$LLT+LTL+TLL$  &  $d-1$ & $d-1$  & $d-1$  &  $0$ & $3(d-1)$\\
\hline
$LLL$   &  $1$ & $0$  & $0$  &  $0$ & $1$\\
\hline
dim   &  $\frac{d(d+1)(d+2)}{6}$ & $\frac{d(d^2-1)}{3}$  & $\frac{d(d^2-1)}{3}$  &  $\frac{d(d-1)(d-2)}{6}$ & $d^3$\\
\hline
\end{tabular}
\end{center}
\caption{Dimensions of projected spaces in $d$ dimensions.}
\label{t1}
\end{table}

All of these spaces are representations of $SO(d-1)$,
some irreducible and others not.
In order to obtain the irreps, let us note that the
$hs$- and $ha$-projections of $\frac32 LTT$ and
$TTL+TLT-\frac12LTT$ are themselves projectors.
Finally, in several of these representations one can
isolate the ``trace'' and the ``tracefree'' part.
In dimension $d=4$, the  $SO(3)$  irreducible representations are then given in Table \ref{t2}, together with the spin and parity carried by them.
For completeness we also list the representations carried
by the two-index symmetric tensor $h$.
The subscripts refer to the number in the labelling of the projectors.

\bigskip
\begin{table}
\begin{center}
\begin{tabular}{|c|c|c|c|c|}
\hline
   & $ts$ & $hs$ & $ha$ & $ta$ \\
\hline
$TTT$ & $3^-$, $1^-_1$ & $2^-_1$, $1^-_2$  & $2^-_2$, $1^-_3$ & $0^-$ \\
\hline
$TTL+TLT+LTT$  & $2^+_1$, $0^+_1$ & - & -  & $1^+_3$  \\
\hline
$\frac32 LTT$   & - & $2^+_2$, $0^+_2$  & $1^+_2$, & - \\
\hline
$TTL+TLT- \frac12 LTT$ & - & $1^+_1$ & $2^+_3$, $0^+_3$  & -  \\
\hline
$TLL+LTL+LLT$   & $1^-_4$ & $1^-_5$  & $1^-_6$   &  - \\
\hline
$LLL$    & $0^+_4$ & -  & -  &  - \\
\hline
\end{tabular}
\qquad
\begin{tabular}{|c|c|}
\hline
 & $s$ \\
\hline
$TT$ & $2^+_4$, $0^+_5$ \\
\hline
$TL$  & $1^-_7$\\
\hline
$LL$   &  $0^+_6$ \\
\hline
\end{tabular}
\end{center}
\caption{$SO(3)$ spin content of projection operators for $A$ and $h$ in $d=4$. ($ts/ta$=totally (anti)symmetric; $hs/ha$=hook (anti)symmetric}
\label{t2}
\end{table}

A given representation of the group $SO(3)$ may appear
more than once in the decomposition of $A_{cab}$.
These copies will be distinguished by a label $i$.
Thus for example the representation $2^-$ occurs twice,
and the two instances are denoted $2^-_1$ and $2^-_2$.
In addition, the same representation may occur also in the
decomposition of the two-tensor $h_{ab}$.
We use the same label for all these representations.
Thus for example the representation $2^+$ occurs altogether
four times: the representations $2^+_i$ with $i=1,2,3$
come from $A_{cab}$ whereas $2^+_4$ comes from $h_{ab}$.

For each representation $J^\cP_i$ there is a
projector denoted $P_{ii}(J^\cP)$.
In addition, for each pair of representations with the
same spin-parity, labelled by $i$, $j$, there is an
intertwining operator $P_{ij}(J^\cP)$.
We collectively refer to all the projectors
and intertwiners as the ``spin-projectors''.
Formulas for all the spin projectors are given in Appendix \ref{proj}.
For convenience they are also given in an ancillary Mathematica notebook on the {\tt arXiv}.

Let us emphasize again that these spin projectors are suitable
to decompose tensors that either have no symmetry property or
are (anti)symmetric in the last two indices.
If one is interested in tensors that are (anti)symmetric in 
the first and third index,
it is more convenient to work with another set of
spin projectors $P'_{ij}(J^\cP)$, such that whenever the
representation $i$ or $j$ is carried by a three-index tensor,
the first two indices are permuted.
For example
\be
P'_{11}(2^+)^{cab}{}_{def}=P_{11}(2^+)^{acb}{}_{edf}\ ,\qquad
P'_{14}(2^+)^{cab}{}_{ef}=P_{14}(2^+)^{acb}{}_{ef}\quad\mathrm{etc}\ .
\label{altproj}
\ee
Similarly one can deal with tensors that are (anti)symmetric
in the first two indices.

\bigskip
\begin{table}
\begin{center}
\begin{tabular}{|c|c|c|c|c|}
\hline
 $J^P$  & $A$ & $h$ & \# of irreps & \# of fields \\
\hline
$3^-$ & 1 & -  & 1 & 7 \\
\hline
$2^+$  & 1,2,3 & 4 & 4  & 20  \\
\hline
$2^-$   & 1,2 & - & 2 & 10 \\
\hline
$1^+$ & 1,2,3 & - & 3  & 9  \\
\hline
$1^-$   & 1,2,3,4,5,6 & 7  & 7   & 21 \\
\hline
$0^+$    & 1,2,3,4 & 5,6  & 6  &  6 \\
\hline
$0^-$    & 1 & -  & 1  &  1 \\
\hline
total    &  &   &   &  74 \\
\hline
\end{tabular}
\end{center}
\caption{Count of fields of general MAG: irreps of
given spin contained in $A$ (2nd column) in $h$ (3rd column);
their total number (4th column) and total number of fields they carry in $d=4$.}
\label{t3}
\end{table}

\subsection{Rewriting the quadratic action}
The projector $P_{ij}(J^\cP)$ has two sets of hidden indices:
one for the representation $J^\cP_i$ and one for the representation
$J^\cP_j$.
These multi-indices $A$, $B$... consist of either three or two indices,
depending whether the carrier field of the representation 
is $A$ or $h$.
Thus for example $P_{11}(2^+)$ has indices
$P_{11}(2^+)^{cab}{}_{def}$,
$P_{41}(2^+)$ has indices
$P_{41}(2^+)^{ab}{}_{def}$ etc.
The spin projectors satisfy the orthonormality relation
\be
P_{ij}(J^\cP)^A{}_B P_{kl}(I^{\cal Q})^B{}_C=
\delta_{IJ}\,\delta_{\cal PQ}\,\delta_{jk}\,\ 
P_{il} (J^\cP)^A{}_C\ ,
\label{ortho}
\ee
and the completeness relation
\be
\sum_{J,\cP,i} P_{ii}(J^\cP)={\bf 1}\ .
\label{complete}
\ee

The linearized quadratic action (3.3), can be written as
\be
S^{(2)}=\frac12\int d^dq\ 
\begin{pmatrix} A(-q) & h(-q)\\ \end{pmatrix}
\begin{pmatrix}{\cal O}_{AA}(q) &{\cal O}_{Ah}(q)
\\ {\cal O}_{hA}(q) & {\cal O}_{hh}(q)
\\ \end{pmatrix}
\begin{pmatrix} A(q) 
\\ h(q)\\ 
\end{pmatrix}\ .
\label{S2}
\ee
In four dimensions, the kinetic operator is
an $74\times 74$ matrix, that we have written
as $64\times 64$, $10\times 10$ and off-diagonal
$10\times 64$ and $64\times 10$ blocks.
Since the operator is Lorentz-covariant,
it maps states of a given spin and parity to states
of the same spin and parity.
Therefore, decomposing $A_{cab}$ and $h_{ab}$
into irreducible representations of the rotation group puts the kinetic operator in block diagonal form. 

Expanding the operator ${\cal O}_{AB}$ in terms of these projection operators
one can rewrite the quadratic action as
\be
S^{(2)}= \frac12\int d^d q\sum_{J Pij}\Phi(-q)\cdot
a_{ij}(J^\cP)\,P_{ij}(J^\cP)\cdot \Phi(q)\ ,
\ee
Exploiting the relations (\ref{ortho},\ref{complete}),
the matrix elements $a_{ij}(J^\cP)$,
where both representations $J^\cP_i$ and $J^\cP_j$
are carried by $A$, can be obtained by
\bea
a_{ij}(J^\cP) &=& \frac{1}{d(J^\cP)} P_{ij}(J^\cP)^{cab}{}_{def}
\cO_{AA}^{def}{}_{cab}
\nn\\
&=& \frac{1}{d(J^\cP)}
P_{ki}(J^\cP)^{cab}{}_{def}\,
\cO_{AA}^{def}{}_{\ell mn}\,
P_{jk}(J^\cP)^{\ell mn}{}_{cab}\ ,\qquad   
\eea
for any fixed $k$, and where $d(J^\cP)$ is the dimension
of the representation $J^\cP$.
The second equality follows from \eq{ortho}, and it shows that it suffices to know the projections operators $P_{jk}$ for any fixed $k$ in order to obtain all coefficients matrices. This was also observed in \cite{baikov}, where $P_{jk}$ for a fixed $k$ ( chosen for convenience to give the simplest projector) were referred to as `semi-projectors'.
Similarly, if the representation $J^\cP_i$ is carried by $A$
and $J^\cP_j$ is carried by $h$, we can use for example
\bea
a_{ij}(J^\cP) &=& \frac{1}{d(J^\cP)} P_{ij}(J^\cP)^{de}{}_{cab}\,
\cO_{Ah}^{cab}{}_{de}
\nn\\
&=& \frac{1}{d(J^\cP)} P_{ki}(J^\cP)^{cab}{}_{def}\,
\cO_{Ah}^{def}{}_{mn}\, P_{jk}(J^\cP)^{mn}{}_{cab}\ ,
\eea
where we have chosen $k$ that is carried by $A$.
These matrices $a_{ij}(J^\cP)$ will be referred to
as the ``coefficient matrices''.
For a general MAG in four dimensions, they are given
in Appendix B.1.

\section{Constraints for ghost- and tachyon-freedom}
\noindent

Let us arrange the fluctuations into a multi-field $\Phi_A$
and introduce corresponding sources:
\be
\Phi_A = \begin{pmatrix} A_{cab}\\ h_{ab}\end{pmatrix}\ ,\qquad {\cal J}_A = \begin{pmatrix} \tau_{cab}\\ \sigma_{ab}\end{pmatrix}\ .
\ee
Adding source terms, the linearized action can be written
\be
S^{(2)}=\int d^d q\ \left[\frac12 \sum_{J Pij}\Phi(-q)\cdot
a_{ij}(J^\cP)\,P_{ij}(J^\cP)\cdot \Phi(q) + {\cal J}(-q)\cdot \Phi(q)\right]\ ,
\label{linactsource}
\ee
which gives  the field equations
\be
\sum_{J Pij} a_{ij}(J^\cP)\,P_{ij}(J^\cP)\cdot \Phi = -{\cal J}\ .
\label{fe1}
\ee
Inverting for $\Phi$ as a function of ${\cal J}$ and
substituting back into $S^{(2)}$ we obtain a quadratic form in ${\cal J}$
that we identify with the saturated propagator and we denote by $\Pi$.
There is, however, a complication:
in a given spin-parity sector, the matrix $a_{ij}$ may have null eigenvectors. This corresponds to the presence of gauge symmetries as follows. Suppose for a given $J^P$, the matrix $a_{ij}$ is $n\times n$ and has rank $m$, thereby admitting $(n-m)$ null vectors,
\be
\sum_{j} a_{ij} V_j^{(r)} = 0\ ,\qquad i,j=1,...,n, \quad r=1,...,n-m\ ,
\ee
Then \eqref{linactsource}
is easily seen to be invariant under
\be
\delta \Phi = \sum_{k,r} V_k^{(r)} P_{k\ell}\cdot \xi^{(r)}\ ,
\qquad \forall\, \ell\ ,\label{gs}
\ee
where $\xi^{(r)}$ are arbitrary functions of the coordinates, 
provided that the sources obey the constraints
\be
\sum_i V_i^{\dagger (r)} P_{ji} \cdot {\cal J} =0\ ,\qquad \forall\, j,r,J,{\cal P}\ ,
\label{sc}
\ee
The preceding analysis has to be repeated in each spin sector
to determine all the gauge symmetries and source constraints.
In practice this cumbersome procedure will not be necessary for the following reasons. 

Let us distinguish gauge symmetries that are already present
in the original action \eqref{action} from ``accidental''
symmetries that are only present in the linearized action.
The latter are broken by interactions and therefore
cannot be maintained in the quantum theory.
In the following we shall restrict ourselves to theories
that do not have accidental symmetries.
Thus, the only infinitesimal gauge invariance
is given by the diffeomorphisms \eqref{lindiff2}:
\be
\delta A_{cab}=-q_b q_c\xi_a\ ,\qquad
\delta h_{ab}=i(q_a\xi_b+q_b\xi_a)\ .
\label{gts}
\ee
Writing this schematically as $\delta\Phi=D\xi$,
since $D\xi$ is a null eigenvector of the linearized kinetic term,
we must have
\be
\sum_{J Pij}a_{ij}(J^\cP)\,P_{ij}(J^\cP)D\xi=0\ .
\ee
Explicit calculation shows that $P_{ij}(J^\cP)D\xi$ is only nonzero
for $J^\cP=1^-$ and $j=4,5,6,7$ or $J^\cP=0^+$ and $j=4,6$.
Then one finds that $a(1^-)$ has the null eigenvector
\be
(0,0,0,-i|q|/\sqrt6, -i|q|/(2\sqrt3), i|q|/2, 1)\ ,
\ee
and $a(0^+)$ has the null eigenvector
$(0,0,0,i|q|/2,0,1)$.
Thus, in general, the ranks of the coefficient matrices $a(1^-)$ and $a(0^+)$ are 6 and 5, respectively.
Invariance of the source term then demands that the sources satisfy the constraint
\footnote{In the tetrad formulation of the theory, the antisymmetric part of the tetrad fluctuation transform as $\delta h_{[ab]}= -\lambda_{ab} + \partial_{[a} \xi_{b]}$, where $\lambda_{ab}$ is the local Lorentz parameter. Maintaining the gauge choice $h_{[ab]}=0$ fixes $\lambda_{ab}= \partial_{[a}\xi_{b]}$. Since $\delta A_{cab}=\partial_c \lambda_{ab}$, one finds \eq{gts}, and hence the source constraint \eq{diffsourceconstr}. }
\be
2i q^a\sigma_{ac}+q^a q^b\tau_{bca}=0\ .
\label{diffsourceconstr}
\ee

To obtain the propagator sandwiched between physical sources one takes the inverse of any $m\times m$ submatrix of $a_{ij}$ with nonzero determinant. 
This amounts to fixing the gauge symmetries and it does not effect the form of the physical saturated propagator \cite{Berends:1979rv}.
Denoting this submatrix by $b_{k\ell}, (k,\ell=1,...m)$, the resulting saturated propagator $\Pi$, upon solving for $\Phi$ in terms of the source and substituting back into the action, takes the form
\be
\Pi = -\frac12 \sum_{J,\cP,k,\ell}\, 
b^{-1}_{k\ell}(J^\cP) \, {\cal J}^\dagger\cdot  
P_{k\ell}(J^\cP) \cdot {\cal J}= -\frac12 \sum_{J,\cP,k,\ell}\, 
\frac{1}{\det b(J^\cP)}\, C_{k\ell}(J^\cP)\, {\cal J}^\dagger\cdot  P_{k\ell}(J^\cP) \cdot {\cal J}\ ,
\label{sp}
\ee
where $C_{k\ell}$ is the transpose of the cofactor matrix associated with the matrix $b$, which is assumed to have rank $m$. 
It is important to stress that in our notation $b^{-1}_{k\ell}$
denotes the matrix element of $b^{-1}$ in the representations
$k$, $\ell$, which need not agree with the element of the matrix
$b^{-1}$ in the $k$-th row and $\ell$-th column (unless $a$ is non-degenerate, in which case $b=a$).
Given that $b_{ij}(q)$ is a hermitian matrix and its momentum dependence is polynomial, the poles at non-vanishing values of $q^2$ can only come from $\det b (J^{\cP})$.
We assume that for each given $J^P$ there will be $s$ propagating
particles, with $s\leq m$.
Then we can write
\be
\det\,b =  C(q^2+m_1^2)\cdots (q^2 +m_s^2)\ ,
\label{fac1}
\ee
where $(C,m^2_1,\ldots,m^2_s)$ are constants.
For a physical spectrum these constants must be real and to simplify the analysis we shall further assume that the masses
$m_n^2$, $n=1,...,s$, are nonvanishing and distinct
(possibly, one of the masses could be zero).
The determinant $\det b$ has a simple zero for $q^2=-m_n^2$,
so exactly one eigenvalue of $b$ must have a zero there.
This implies that the residue matrix 
\be
\lim_{q^2\to-m_n^2} (q^2+m_n^2)b^{-1}
\ee
has exactly one non-vanishing eigenvalue. 

Before proceeding to implication of this for ghost-freedom criteria, we need to first note that the spin projectors in (\ref{sp}) contain powers of $1/q^2$ that do
not contribute to the physical propagators.
These spurious poles at zero momentum, which we shall sometimes refer to as kinematical singularities, cancel out in the full saturated propagator. These poles arise from the product of constants, or $1/(q^2+m^2)$, with the longitudinal parts of the spin projection operators. In the latter case, the simple procedure of partial fractions gives rise to terms in which the spin projection operator are evaluated on the mass shell, plus terms with powers of $1/q^2$. 
For example:
\be
\frac{1/q^2}{q^2+m^2} = \frac{1/(-m^2)}{q^2+m^2}+\frac{1/m^2}{q^2}\ ,
\ee
and similarly for expressions of the form $1/((q^2)^n (q^2+m^2))$. The first term on the r.h.s. has the same pole at $q^2=-m^2$, but in its coefficient the momentum squared is now evaluated
at the pole. The second term gives another spurious pole at zero.
In the end all the spurious poles cancel out and we remain with a combination of the spin projectors evaluated on the mass shell or constants sandwiched between sources that obey source constraints. 

With the issue of kinematical singularities out of the way, we can now state the conditions for the absence of ghosts and tachyons. The tachyon-freedom condition is very simple, namely
\be 
\mbox{Tachyon-free} \ \ \implies \ \  m_n^2 > 0\ ,\quad  n=1,...,s\ .
\ee

To examine the ghost-freedom condition, it is convenient to diagonalize the matrix $b^{-1}$ by means of a unitary matrix $U$:
$Ub^{-1}U^\dagger=D$, with
$D_{ij}=\lambda_i\delta_{ij}$.
Then, the saturated propagator is
\begin{align}
\Pi &= -\frac12 \sum_{J,\cP,i,j,k} \lambda_k\,U^\dagger_{ik} U_{kj} \,\cJ^\dagger\cdot P_{ij}\cdot \cJ\ .
\label{sp33}
\end{align}
where we have suppressed the $J^\cP$ labels for brevity in notation. Next, defining
\be
{\widehat \cJ} = U_{ij} P_{ij} \cdot \cJ\ ,
\ee
and using the orthonormality property of the spin projectors, one finds that \cite{Lin:2018awc}
\be
\Pi =-\frac12 \sum_{J,\cP,k} \lambda_k\,{\widehat \cJ}\cdot P_{kk}\cdot{\widehat \cJ}\ .
\label{hj}
\ee
As already remarked, precisely one eigenvalue has nonzero residue at a given pole, and the residue of $\Pi$ at a given pole must be positive for ghost freedom.  Therefore, noting that on-shell 
\be
T_{ab}\Big|_{q^2=-m^2}= \delta_{ai}\delta_{bj}\ ,\qquad L_{ab}\Big|_{q^2=-m^2}=-\delta_{a 0}\delta_{b0}\ ,\quad i=1,2,3\ ,
\label{osp}
\ee
we conclude that the ghost-free condition can be written as\footnote{The sign depends on the signature of the metric. It may be useful to recall that in our signature,
for a massive scalar field, $b=-(q^2+m^2)$.} 
\be
\mbox{Ghost-free}\ \ \implies \ \ \underset{q^2=-m_n^2}{\mbox{Res}}\,
\sum_{k=1}^s \left[b^{-1}_{kk}(J^\cP) (-1)^{n_L(k)+1} \right] >0\ , \quad \forall{J,\cP, n}\ ,
\label{gf11}
\ee
where $n_L(k)$ is the number of $L$ factors in $P_{kk}(J^\cP)$.  This grading is due to the fact that each $L$ factor in the diagonal spin projector gives a minus sign on the mass-shell\footnote{In a convention where $\eta_{ab} ={\rm diag} (+---)$, the signs in \eq{osp} flip, and the grading is given by the number $n_T(k)$ of $T$ factors in $P_{kk}(J^\cP)$. 
Since this number $\mod 2$, is the same in any $J^\cP$ sector, the ghost-free condition becomes  $\underset{q^2=-m_k^2}{\mbox{Res}}
{\rm tr} \left[-b^{-1}(J^\cP) (-1)^P \right] > 0$, in any $J^\cP$ sector, and where $\tr$ is the ordinary trace.}.
\smallskip

Going back to the formula \eq{sp}, or \eq{hj}, in any $\cJ^\cP$ sector involving the matrix $b^{-1}$ with rank greater than one, there will clearly be mixing of sources that survive the source constraints. Given that all the kinematical singularities have cancelled, the result for the saturated propagator in such $J^\cP$ sectors can be written in such a way that the standard form of the spin $J^\cP$ propagators arise in terms of a suitable combination of these sources. This phenomenon will be clearly shown in the multi-parameter models analysed below; see \eq{n2} and \eq{tff}. 

Given any MAG with specific couplings $c_1\ldots c_{16}$, $a_0,a_1\ldots a_{11}$
one can use these conditions on the coefficient matrices given in
Appendix B.1,  and determine the spectrum of the theory.
However, the 28-parameter class of all MAGs is too broad for a general analysis,
so in the following we discuss two important subclasses:
MAGs with either $Q=0$ or $T=0$.

\section{Theories with metric connection}

\subsection{General case}

In metric theories the following identities hold:
\be
Q_{\lambda\mu\nu}=0\ ,\quad
F_{\mu\nu(\rho\sigma)}=0\ ,\quad
F^{(14)}_{\mu\nu}=-F^{(13)}_{\mu\nu}\ ,\quad
F_{\mu\nu}=0\ .
\label{sr}
\ee
Using these properties, the most general action up to and including curvature and torsion squared terms is a 10-parameter action given by 
\bea
S(g,A) &=&-\frac12
\int d^d x\ \sqrt{|g|}\,\Big[-a_0 F+ F^{\mu\nu\rho\sigma} 
\big(g_1 F_{\mu\nu\rho\sigma} 
+g_3 F_{\rho\sigma\mu\nu}
+g_4 F_{\mu\rho\nu\sigma})
\label{actionmetric}\w2
&&  
+ F^{(13)\mu\nu} \big(g_7 F^{(13)}_{\mu\nu} 
+g_8 F^{(13)}_{\nu\mu} \big)
+ g_{16} F^2
+ T^{\mu\rho\nu} \big(b_1 T_{\mu\rho\nu} 
+b_2 T_{\mu\nu\rho}\big)  
+ b_3 T^\mu T_\mu
\Big]\ ,
\nn
\eea

Note that the metricity condition $Q=0$ is a kinematic constraint that changes
the nature of the theory: the action \eqref{actionmetric} is not obtained from the general MAG action \eqref{action} simply by specializing the values of the couplings.
Nevertheless, it is useful to write it in the same form and to
preserve the numbering of the invariants. To distinguish the two cases, we changed the name of the couplings from $c_i$ to $g_i$ and from $a_i$ to $b_i$.
\begin{table}
\begin{center}
\begin{tabular}{|c|c|c|c|c|}
\hline
 $J^P$  & $A$ & $h$ &  \# of irreps & \# of fields \\
\hline
$3^-$ & - & -  & 0 & 0 \\
\hline
$2^+$  & 3 & 4 & 2  & 10  \\
\hline
$2^-$   & 2 & - & 1 & 5 \\
\hline
$1^+$ & 2,3 & - & 2  & 6  \\
\hline
$1^-$   & 3,6 & 7  & 3   & 9 \\
\hline
$0^+$    & 3 & 5,6  & 3  &  3 \\
\hline
$0^-$    & 1 & -  & 1  &  1 \\
\hline
total    &  &   &   &  34 \\
\hline
\end{tabular}
\end{center}
\caption{Count of fields of metric MAG: irreps of
given spin contained in $A$, in $h$,
their total number, and number of fields they carry
in $d=4$.}
\label{t4}
\end{table}
Notwithstanding the fact that the action \eq{actionmetric} is not a special case of \eq{action}, it is possible to linearize it by making use of the results already computed for the general action \eq{action} as follows. Let us first consider the $F^2$ terms. In the action \eq{actionmetric}, and in accordance with \eq{sr}, making the substitutions
\be
F_{abcd}\ \to \ \frac12 \left( F_{abcd}-F_{abdc}\right)\ ,\qquad  F_{ab}^{(13)} \ \to \ \frac12 \left(F_{ab}^{(13)}-F_{ab}^{(14)} \right)\ ,
\ee
and comparing the result with the general action \eq{action}, we obtain the relations
\bea
c_1&=&\frac12 g_1\ ,\quad
c_2=-\frac12g_1\ ,\quad 
c_3=g_3\ ,\quad
c_4=\frac14 g_4\ ,\quad
c_5=-\frac12g_4\ ,\quad
c_6=\frac14 g_4\ ,\quad
\nn\\
c_7&=&\frac14 g_7\ ,\quad 
c_8=\frac14 g_8\ ,\quad
c_9=\frac14 g_7\ ,\quad
c_{10}=\frac14 g_8\ ,\quad
c_{11}=-\frac12 g_7\ ,\quad
c_{12}=-\frac12 g_8\ ,
\nn\\
c_{13}&=&c_{14}=c_{15}=0\phantom{\frac12}\ ,\quad
c_{16}=g_{16}\ .
\label{metricc}
\eea

Next, let us consider the substitution required for the parameters $a_i$ in terms of $b_i$. This is more subtle due to the fact that expanding around $A_{cab}=0$, the variation of the metricity condition implies that the fluctuation fields are related by
\be
\partial_c h_{ab}= A_{cab}+ A_{cba}\ ,
\label{metfluc}
\ee
where we recall that $A$ denotes also the fluctuation. Thus inserting in the linearized action the decomposition  $A_{cab}= A_{c[ab]} + A_{c(ab)}$, 
the symmetric part of $A$ gives terms proportional to $h$ that can be compared to those that, in a general MAG, are produced by $Q$.
This gives the relations
\be
a_1=b_1\ ,\quad 
a_2=b_2\ ,\quad 
a_3=b_3\ ,\quad
a_4=-a_5= \frac12 b_1+\frac14b_2\ ,
\nn
\ee
\be
a_6=a_7=\frac14b_3\ ,\quad
a_8=-\frac12b_3\ ,\quad
a_9=2b_1+b_2\ ,\quad
a_{10} =-a_{11}=-b_3\ .
\label{metrica}
\ee
In summary, the coefficient matrices of the metric theory are obtained
from those of the general MAG by inserting the values for the couplings
$c_i, a_i$ in terms of $g_i, b_i$ as given in \eq{metricc} and \eq{metrica}, and deleting all the rows and columns that pertain
to representations carried by symmetric three-tensors. 
The remaining representations, and the count of degrees of freedom
that they carry,  is given in Table 4.
The coefficient matrices of metric MAG in $d=4$ 
are given explicitly in Appendix B.2.

\subsection{Neville's model}

In order to test of our formulae and procedures we reconsider here,
as an example,  
the Neville model \cite{neville}, which is the same as model (ii) in  \cite{Sezgin:1979zf}. 
It corresponds to choosing the couplings
$g_1=g_3=-g_4/4 ~\equiv~ g$, $g_7=g_8=g_{16}=0$, $b_1=b_2=b_3=0$.

In the sectors $1^-$ and $0^+$,
to fix diffeomorphism invariance,
we choose the non-degenerate $b$-matrices to be the upper left $2\times 2$
sub-matrices of the general $a$-matrices given in Appendix B.2, namely
$b^{-1}_{ij}(1^-)$ with $i,j=3,6$ and
$b^{-1}_{ij}(0^+)$ with $i,j=3,5$.
The inverses of these coefficient matrices are then given by:
\be
b^{-1}(2^+)=\frac{1}{a_0}
\begin{pmatrix} 
0 &  \frac{2i\sqrt2}{|q|}
\\
- \frac{2i\sqrt2}{|q|} & -\frac{4}{q^2}
\end{pmatrix}
\ ,\quad
b^{-1}(2^-)=\frac{2}{a_0}\ ,
\quad
b^{-1}(1^+)=\frac{1}{a_0}
\begin{pmatrix} 2 &  0
\\
0 &-1
\end{pmatrix}\ ,
\ee

\be
b^{-1}(1^-)=\frac{1}{a_0} 
\begin{pmatrix} 
0 &  -\sqrt2
\\
- \sqrt2 &1
\end{pmatrix}
\ ,\quad
b^{-1}(0^+)=\frac{1}{a_0} 
\begin{pmatrix} 
0 & - \frac{i\sqrt2}{|q|}
\\
\frac{i\sqrt2}{|q|} &\frac{2}{q^2}
\end{pmatrix}
\ ,\quad
b^{-1}(0^-)=-\frac{1}{a_0+6g q^2}\ .
\nn
\ee
The analysis of section 4.4 shows that this theory contains a 
massless graviton and a massless pseudoscalar state,
with mass $m^2=a_0/(6g)$.
Absence of tachyons and ghosts requires
$a_0>0$ and $g>0$. The saturated propagator is:
\bea
\Pi\!\!\!&=&\!\!\!-\frac12\int d^4q\Biggl\{
{\cal J}
\left(\sum_{i,j=3,4} b_{ij}^{-1}(2^+)P_{ij}(2^+)
+\sum_{i,j=3,5} b_{ij}^{-1}(0^+)P_{ij}(0^+)\right){\cal J}
\\
&&
\!\!\!\!\!\!
+\tau\,\Big[b^{-1}(2^-)\,P_{22}(2^-)
+\sum_{i,j=2,3} b_{ij}^{-1}(1^+)P_{ij}(1^+)
+\sum_{i,j=3,6} b_{ij}^{-1}(1^-)P_{ij}(1^-)
+b^{-1}(0^-)\,P(0^-)
\Big]\,\tau
\Biggr\}
\nn
\eea

As discussed in section 4.4, and using the source constraint \eq{diffsourceconstr}, it can be rewritten in a more explicit form, where the spin projection operators are put on shell:
\bea
\Pi &=& -\frac{1}{2a_0}
\int dq\, \Bigg\{\tau\cdot\left(-\frac{m^2}{q^2+m^2} P(0^-,m^2) 
+2 P_{22}(2^-,\eta) 
\right)\cdot\tau
\nn\\
&&
\qquad\qquad\qquad
-\frac{4}{q^2} 
S\cdot
\left(P_{44}(2^+,\eta)-\frac12 P_{55}(0^+,\eta) \right)
\cdot S\Bigg\}
\ ,
\label{n2}
\eea
where
$S_{ab}=\sigma_{ab}+iq^c\tau_{acb}\sigma_{ab}$,
and following \cite{Sezgin:1979zf} we have used
\be
P(J^\cP,m^2) \equiv P(J^\cP,q)\Big|_{q^2=-m^2}\ .
\ee
\be
P(J^\cP,\eta) \equiv P(J^\cP,q)\Big\vert_{\partial\to 0}\ .
\ee
The last term is the standard graviton propagator
\be
\int dq\, S^{ab} (-q)~ \frac{2}{a_0 q^2}\left( \eta_{ac}\eta_{bd}-\frac12 \eta_{ab}\eta_{cd}\right) S^{cd}(q)\ ,
\ee
%
%
while for the spin $0^-$ we have
\be
\tau\cdot P(0^-,m^2)\cdot\tau
=\tau^{[cab]}\left(\eta_{cd}+\frac{q_c q_d}{m^2}\right)
\left(\eta_{ae}+\frac{q_a q_e}{m^2}\right)
\left(\eta_{bf}+\frac{q_b q_f}{m^2}\right)\tau^{[def]}\ .
\ee 
The spin $1^+$ and $1^-$ contributions actually vanish.

\section{Torsion-free theories}

\subsection{General case}

In torsion-free theories the following identities hold:
\be
T_\mu{}^\rho{}_\nu =0\ ,\quad
F_{[\mu\nu}{}^\rho{}_{\sigma]}=0\ ,
\quad F_{\mu\nu}= -2F^{(13)}_{[\mu\nu]}\ .
\label{vt}
\ee
These reduce the number of independent invariants.
One finds that the terms in \eqref{action} 
with parameters $c_5, c_6,c_{13}, c_{14},c_{15},a_1,a_2,a_3,a_9,a_{10},a_{11}$ become redundant.
Thus we parameterize the most general torsion-free MAG action as
\bea
S(g,A) &=&-\frac12\int d^dx\ \sqrt{|g|}\,\Big[ -a_0 F+ F^{\mu\nu\rho\sigma} \big( h_1 F_{\mu\nu\rho\sigma} 
+ h_2 F_{\mu\nu\sigma\rho} 
+ h_3 F_{\rho\sigma\mu\nu} 
+ h_4 F_{\mu\rho\nu\sigma} \big)
\nn\w2
&&  
\qquad\qquad
+ F^{(13)\mu\nu} \big(h_7 F^{(13)}_{\mu\nu} + h_8 F^{(13)}_{\nu\mu} \big)
+ F^{(14)\mu\nu} \big( h_9 F^{(14)}_{\mu\nu} 
+ h_{10} F^{(14)}_{\nu\mu}\big) 
\nn\w2
&& 
\qquad\qquad
+ F^{(14)\mu\nu}\big(h_{11} F^{(13)}_{\mu\nu}
+ h_{12} F^{(13)}_{\nu\mu} \big)
+h_{16}F^2
\nn\w2
&& 
\qquad\qquad
+Q^{\rho\mu\nu}\big( a_4 Q_{\rho\mu\nu} 
+ a_5 Q_{\nu\mu\rho}\big)  
+ a_6 Q^\mu Q_\mu + a_7 \tQ^\mu \tQ_\mu + a_8 Q^\mu \tQ_\mu  
\Big]\ ,
\label{actiontf}
\eea
Once again we note that $T=0$ is a kinematic constraint,
so that the theories we now consider are not equivalent to just
setting to zero the parameters listed above.
For this reason, the remaining parameters $c_i$ have been renamed $h_i$.

In the torsion-free case, the field $A_{\lambda\mu\nu}$ 
is symmetric in $\lambda$, $\nu$.
In four dimensions, this reduces the number of degrees of freedom of $A$ from 64 to 40.
The corresponding spin representations are listed in the second column
of Table 5.
In order to obtain the coefficient matrices, we use the ``primed'' spin projectors defined in the end of section 3.2, which are better suited to decompose a tensor symmetric in the first and last index.
All the primed spin projectors in the columns
$ha$ and $ta$ in Table 2 give zero when acting on a torsion-free connection.
Thus, the coefficient matrices for this case are smaller:
their dimensions are given by the fourth column of Table 4.
A diffeomorphism \eqref{diffeo} preserves the symmetry
of $A_{\lambda\mu\nu}$ and 
diffeomorphism symmetry reduces by one the rank of
the coefficient matrices for spins $1^-$ and $0^+$.
The coefficient matrices for the torsion-free theory 
in four dimensions are given in Appendix B.3.

\bigskip
\begin{table}
\begin{center}
\begin{tabular}{|c|c|c|c|c|}
\hline
 $J^P$  & $A$ & $h$ & \# of irreps & \# of fields \\
\hline
$3^-$ & 1 & -  & 1 & 7 \\
\hline
$2^+$  & 1,2 & 4 & 3  & 15  \\
\hline
$2^-$   & 1 & - & 1 & 5 \\
\hline
$1^+$ & 1 & - & 1  & 3  \\
\hline
$1^-$   & 1,2,4,5 & 7  & 5   & 15 \\
\hline
$0^+$    & 1,2,4 & 5,6  & 5  &  5 \\
\hline
$0^-$    & - & -  & 0  &  0 \\
\hline
total    &  &   &   &  50 \\
\hline
\end{tabular}
\end{center}
\caption{Count of fields of torsion-free MAG: irreps of
given spin contained in $A$, in $h$,
their total number, and number of fields they carry
in $d=4$.}
\label{t44}
\end{table}

\subsection{Torsion-free theories with projective symmetry}

Let us now examine the possible additional symmetries in this case. 
We find that while the symmetry \eq{pt2} is still too restrictive, in the sense
that it requires all $c$-coefficients to vanish, we can achieve
projective symmetry, which is now a symmetric combination of 
\eq{pt} and \eq{pt3}:
\be
\delta_4 A_\mu{}^\rho{}_\nu = 2 \lambda_{(\mu} \delta_{\nu)}^\rho\ ,\qquad \delta g_{\mu\nu}=0\ .
\label{pt4}
\ee
It follows that
\bea
\delta_4 F_{\mu\nu\rho\sigma} &=& 2g_{\rho\sigma} \nabla_{[\mu}\lambda_{\nu]}-2 g_{\rho[\mu} \nabla_{\nu]} \lambda_\sigma \ ,
\nn\\
\delta_4 Q_{\rho\mu\nu} &=& 2\lambda_\rho g_{\mu\nu}+2g_{\rho(\mu} \lambda_{\nu)}\ .
\eea
Invariance of the action is found to require that 
\bea
h_1 &=& \frac14\left[ -2h_7+2dh_8+(d-1)h_{11} + (d+2)h_{12}+2(1-d)h_{16}\right]\ ,
\nn\\
h_3&=&  \left[2 h_2 + \frac{d}{2} \left(h_{11} + h_{12}\right) +(1-d)h_{16}\,\right]\ ,
\nn\\
h_4 &=& \frac12\left[-4h_2 + 2(2-d)h_7 + 2(1-2d)h_8  - 2d h_{11}  
- (2d+3) h_{12} + 4(d-1) h_{16}\,\right]\ ,
\nn\\
h_9 &=&  \frac16 \Big[ 2(d-2)h_7 +2(2d-1)h_8 +(d+1)h_{11}  + (2d+5) h_{12} + 6(1-d) h_{16}\Big]\ ,
\nn\\
h_{10} &=& \frac16\Big[ -2(d-2)h_7 -2(2d-1)h_8 -(d-2)h_{11} - 2(d+1) h_{12}
    \Big]\ ,
    \nn\\
a_4 &=&  \frac1{16} \Big[5(1-d)a_0 - 24 (d+1)a_6  + 4 (d+3)a_7 - 2 (d+7) a_8 \Big]\ ,
\nn\\
a_5 &=&  \frac18 \Big[ 3(d-1)a_0 + 8 (d+1) a_6  - 4 (d+3)a_7+2(1-d) a_8  \Big]\ ,
\label{tfp}
\eea
where we have used \eq{vt} and the following formula
\be
\delta_4 \int d^d x\, \sqrt{-g}\, F = \int  d^d x\,\sqrt{-g}\, (1-d) \left(\frac12 Q_\mu -{\widetilde Q}_\mu\right) \lambda^\mu\ ,
\ee
with a total derivative term discarded. 
The part of the action proportional to $h_2$, vanishes due to the identity  
\be
F^{\mu\nu\rho\sigma} \left( F_{\mu\nu\sigma\rho}+2F_{\rho\sigma\mu\nu}-2F_{\mu\rho\nu\sigma} \right) =0\ , 
\ee
which follows from repeated use of the second equation in \eq{vt}. Therefore, the action depends on nine parameters, namely, $(a_0,a_6,a_7,a_8)$ and $(h_7,h_8,h_{11}, h_{12},h_{16})$, and it takes the form
\bea
S(g,A) &=&  -\frac12\int d^d x\ \sqrt{|g|}\,\Big\{ -a_0 F + F^{\mu\nu\rho\sigma}\Big( \gamma_1  F_{\mu\nu\rho\sigma} 
+ \gamma_2 F_{\mu\rho\nu\sigma} 
+\gamma_3 F_{\rho\sigma\mu\nu}\Big) 
+h_{16}F^2
\nn\w2
&&  + F^{(13)\mu\nu} \Big( h_{7} F^{(13)}_{\mu\nu} 
+ h_8 F^{(13)}_{\nu\mu} 
+ h_{11} F^{(14)}_{\mu\nu} 
+ h_{12} F^{(14)}_{\nu\mu}\Big)
+ F^{(14)\mu\nu} \Big( \gamma_4 F^{(14)}_{\mu\nu} 
+ \gamma_5 F^{(14)}_{\nu\mu}\Big)
\nn\w2
&& + Q^{\rho\mu\nu} \Big( \gamma_6 Q_{\rho\mu\nu} 
+ \gamma_7 Q_{\nu\mu\rho}\Big) + a_6 Q_\mu Q^\mu 
+a_7 {\widetilde Q}^\mu {\widetilde Q}_\mu 
+ a_8 Q^\mu {\widetilde Q}_\mu \Big)\ ,
\label{tfpa}
\eea
where the parameters $(\gamma_1,...,\gamma_7)$ are defined in terms of the 9 parameters of the action as  
\bea
\gamma_1 &=& -\frac12 h_7+\frac{d}{2} h_8 +\frac{d-1}{4} h_{11} + \frac{d+2}{4} h_{12} +\frac{1-d}{2} h_{16}\ ,
\nn\\
\gamma_2 &=& (2-d) h_7 +(1-2d) h_8 -d\,h_{11} -\frac{2d+3}{2} h_{12} +2(d-1) h_{16}\ , 
\nn\\
\gamma_3 &=& \frac{d}{2} (h_{11} + h_{12}) +(1-d) h_{16}\ ,
\nn\\
\gamma_4 &=& \frac{d-2}{3} h_7 +\frac{2d-1}{3} h_8 +\frac{d+1}{6} h_{11} + \frac{2d+5}{6} h_{12} +(1-d) h_{16}\ ,
\nn\\
\gamma_5 &=& \frac{2-d}{3} h_7 + \frac{1-2d}{3} h_8 +\frac{2-d}{6} h_{11}-\frac{d+1}{3} h_{12}\ ,
\nn\\
\gamma_6 &=& \frac{5(1-d)}{16} a_0 -\frac{3(d+1)}{2} a_6 +\frac{d+3}{4} a_7 -\frac{d+7}{8} a_8
\nn\\
\gamma_7 &=& \frac{3(d-1)}{8} a_0 +(d+1) a_6 -\frac{d+3}{2} a_7 +\frac{1-d}{4} a_8\ .
\label{tfpp}
\eea
In four dimensions, the projective symmetry eliminates four fields, reducing by one the ranks of the coefficient matrices
$1^-$ and $0^+$. In fact one finds that $a(1^-)$ has the null eigenvectors
\be
\left(\sqrt{10/3}\,i|q|,\sqrt{2/3}\,i|q|,\sqrt{3/2}\,i|q|,0,1\right)\ ,\qquad
\left(\sqrt{10},\sqrt2,\sqrt2,1,0\right)\ ,
\ee
while $a(0^+)$ has the null eigenvectors
\be
\left(-(1/2)\,i|q|,(1/2\sqrt2)\,i|q|,0,0,1\right)\ ,\qquad
\left(1,-1/\sqrt2,1,0,0\right)\ .
\ee
The ranks of the coefficient matrices for the representations
$3^-$, $2^+$, $2^-$, $1^+$, $1^-$, $0^+$ are 1, 3, 1, 1, 3, 3 respectively.

Invariance of the source term
implies that the sources must obey the constraints:
\be
\tau^\nu{}_{\nu\mu}=0\ ,\qquad
\tau_{\mu\nu}{}^\nu=0\ .
\ee
Next, we examine the spectrum of this 9-parameter model.

\subsection{New ghost- and tachyon-free theories}

To further simplify matters, we shall restrict our attention to choices of parameters such that:
\begin{itemize}
\item[(i)]  Spin 3 field does not propagate, and
\item[(ii)] in the spin $2^+$ sector only the massless graviton propagates.
\end{itemize}
Condition $(i)$ is achieved by setting to zero
the coefficient of $-q^2$ in (\ref{sp3tfp}):
\be
h_{16} = \frac16 ( 6 h_7 + 6 h_8 + 5 h_{11} + 5 h_{12} )\ .
\label{no3}
\ee
In order to impose $(ii)$,
we consider the rank 3 matrix $b_{ij}(2^+)=a_{ij}(2^+)$ 
with $i,j=1,2,4$.
Demanding that the determinant of this matrix contains no powers of $-q^2$ higher than one, leads to
\be
h_{12}=-h_{11}\ ,\qquad h_8=-h_7\ .
\label{graviton}
\ee
With these conditions, the class of actions that we consider are of the form
\bea
&& S(g,A)= -\frac12\int d^n x\ \sqrt{|g|}\,\Bigg\{ -a_0 F
-\frac14(10h_7+3h_{11})F^{\mu\nu\rho\sigma}
\Big(F_{\mu\nu\rho\sigma}-2F_{\mu\rho\nu\sigma}\Big)
\nn\\
&&
\qquad\qquad\qquad
+ 2F^{(13)}_{[\mu\nu]}\Big( h_7 F^{(13)\mu\nu} +h_{11}F^{(14)\mu\nu} \Big)
-\frac23(5h_7+4h_{11})\, F^{(14)}_{[\mu\nu]}F^{(14)\mu\nu}
\nn\\
&&
\qquad\qquad\qquad
+\frac{1}{48}(12a_0+A-16B)\,Q^{\rho\mu\nu}Q_{\rho\mu\nu}
-\frac{1}{24}(12a_0-A-8B)\,Q^{\rho\mu\nu}Q_{\nu\mu\rho} 
\nn\\
&&
\qquad\qquad\qquad
-\frac{1}{288}(72a_0+A-32B+49C)\,Q_\mu Q^\mu
-\frac{1}{72}(A-8B+25C)\,\wt Q_\mu\wt Q^\mu
\nn\\
&&\qquad\qquad\qquad
+\frac{1}{72}(36a_0-A-16B+35C)\,Q_\mu\wt Q^\mu
\Bigg\}\ ,
\label{faction}
\eea
where we introduced the following convenient new combinations of parameters:
\bea
A&=& 7 a_0- 40 a_6 - 28 a_7 - 34 a_8\ ,
\nn\\
B&=&4\,a_0 + 20\,a_6 - 7\,a_7 +  2\,a_8\ , 
\nn\\
C &=& a_0 + 8\,a_6 - 4\,a_7 + 2\,a_8 \ .
\eea

Let us now discuss the dynamical content of this theory. We have already ruled out the propagation of a spin-3 state, for which
\be
a(3^-)=-\frac{A}{4}\ .
\ee
In the spin-$2^+$ sector we have
\be
\det b(2^+) =  \frac{1}{32}a_0\,AB\,q^2 \ .
\ee
As is well known, the propagation of a massless spin $2^+$ state
requires an admixture of a spin $0^+$ state.
Having imposed \eq{no3} and \eq{graviton}, 
and fixing the diffeomorphism- and projective-gauge by choosing the non-degenerate coefficient submatrix to be
$b_{ij}(0^+)$ with $i,j=3,4,5$, we get
\be
\det b\,(0^+) = -\frac{1}{16} a_0\, AC\,q^2\ .
\ee
Thus, the existence of a massless graviton requires that $A$, $B$, $C$
and $a_0$ are all nonvanishing. In particular, this implies that the coefficient matrix for the spin 3 sector is not zero.  

As we shall now see, having imposed 
\eq{no3} and \eq{graviton} we find that all the coefficient matrices have maximum rank submatrices whose determinants are at most first order in $q^2$. This means that in any given sector at most one state propagates. 
Indeed, denoting $b(2^-)=a(2^-)_{11}$, $b(1^+)=a(1^+)_{11}$, 
and taking the non-degenerate submatrix 
$b_{ij}(1^-)$  with $i,j=2,4,5$, we find
\bea 
b(2^-) &=& \frac{1}{4} \left[ 2 B+ ( 30 h_7 + 9 h_{11} )\, q^2 \right]\ .
\label{b2m}\w2
b(1^+) &=& \frac{1}{6}  \left[ 3 B + ( 40 h_7+17 h_{11} )\, q^2 \right]\ ,
\label{b1p}\w2
\det b\,(1^-) &= &  -\frac{5A}{288}\Delta\ ,\quad
\label{detb1m}
\mathrm{where}\quad
\Delta = 6BC+ (16B+25C)(2h_7+h_{11}) q^2 \ .
\label{b1m}
\eea
Note that since $A$, $B$, $C$ are nonvanishing,
there is no room for accidental symmetries.
From these equations we can read off the masses of the modes $2^-$, $1^+$ and $1^-$.

We can now list the matrices $b^{-1}_{ij}(J^\cP)$:
\be
b^{-1}(3^-) = -\frac{4}{A}\ ,
\ee
\bea
b^{-1}(2^+)&=&\frac{1}{a_0 q^2} 
\begin{pmatrix} 
-\frac13 q^2 & \frac{2\sqrt 2}{3} q^2  & -\frac{2}{\sqrt 3} i|q|
\\
\frac{2\sqrt 2}{3} q^2 & -\frac{8}{3} q^2 & 4\sqrt{\frac23} i|q|
\\
\frac{2}{\sqrt 3} i|q|& -4\sqrt{\frac23} i|q|& -4
\end{pmatrix} 
+ 
\begin{pmatrix}
-\frac{4}{A}  & 0& 0
\\
0&\frac{2}{B} & 0
\\
0 &  0 & 0
\end{pmatrix}\ ,
\label{binv2p}
\\
b^{-1}(0^+) &=& \frac{1}{a_0 q^2} 
\begin{pmatrix} 
\frac34 q^2 & \frac{1}{2\sqrt2} q^2  & -{\sqrt{3/2}} i|q|
\\
\frac{1}{2\sqrt2} q^2 & \frac16 q^2 & -\frac{1}{\sqrt 3} i|q|
\\
{\sqrt{3/2}} i|q|  &  \frac{1}{\sqrt 3} i|q| & 2
\end{pmatrix}
+  
\begin{pmatrix} 
\frac{1}{C}-\frac{1}{A} & \frac{2\sqrt2}{A} & 0
\\
\frac{2\sqrt2}{A} & - \frac{8}{A} & 0 \\  0 & 0  & 0
\end{pmatrix}\ .
\label{binv0p}
\eea
\bea
b^{-1} (2^-) &=&  \frac{4}{  2 B+ ( 30 h_7 + 9 h_{11} )\, q^2 }\ ,
\label{binv2m}
\w2
b^{-1}(1^+) &=&  \frac{6}{3 B + ( 40 h_7+17 h_{11} )\, q^2 }\ ,
\label{binv1p}
\w2
b^{-1}(1^-) &=&\frac{1}{\Delta} 
\begin{pmatrix} 
4(B+C)+\frac23(2h_7+h_{11})q^2 & 0& \frac{\sqrt2}{3}(6B-12C+13(2h_7+h_{11})q^2)
 \\ 0 & 0 & 0 
 \\ 
\frac{\sqrt2}{3}(6B-12C+13(2h_7+h_{11})q^2)  & 0 & 2B+8C+\frac{169}{3}(2h_7+h_{11})q^2
\end{pmatrix} 
\nn\\&& 
- \frac{1}{15A}
\begin{pmatrix}
2 & 12 & \sqrt2
\\
12 & 72 & 6\sqrt2
\\ 
\sqrt2 & 6\sqrt2 & 1
\end{pmatrix}\ .
\label{binv1m}
\eea

We can now state the ghost- and tachyon-free conditions.
The spin $2^+$ and $0^+$ give the standard
graviton propagator.
In the remaining sectors $2^-$, $1^+$ and $1^-$ we have the positivity conditions for the squared masses
\be
m_2^2=\frac{2B}{3(3h_{11}+10h_7)}>0
\ ,\quad
m_+^2=\frac{3B}{17h_{11}+40h_7}>0
\ ,\quad
m_-^2=\frac{6BC}{(16B+25C)(h_{11}+2h_7)}>0
\label{notac}
\ee
while the corresponding residues have to satisfy the following requirements from (\ref{gf11}):
\be
r_2=\frac{4}{3(3h_{11}+10h_7)}<0
\ ,\quad
r_+=\frac{6}{17h_{11}+40h_7}>0
\ ,\quad
r_-=\frac{12(8B^2+25C^2)}{(16B+25C)^2(h_{11}+2h_7)}<0
\label{nogh}
\ .
\ee
The different sign of $r_+$ is due to the fact that all $P_{kk}(1^+)$ contain one $L$ factor. 
One easily sees that in general these conditions cannot be fulfilled simultaneously: taken together, the conditions for the spin $2^+$ imply $B<0$ while those for the spin $1^+$ imply $B>0$.

The only way to find consistent propagation then is to suppress the propagation of either one of these two degrees of freedom.
If we set $h_{11}=-\tfrac{10}{3}h_7$, the mass $m_2^2$ becomes infinite.
In this case we remain with
\be
r_+=-\frac{9}{25h_7}>0
\ ,\quad
r_-=-\frac{7(8B^2+25C^2)}{(16B+25C)^2h_7}<0
\ .
\ee
that, taken together, are contradictory for generic $B$ and $C$.
However, we can further suppress the propagation of the $1^-$ state by setting 
$C=-16B/25$ (in this case $\det b(1^-)$ becomes independent of $q^2$)
and we remain just with the massless graviton and a massive $1^+$ state 
with squared mass $m_+^2=-\frac{9B}{50h_7}$,

Similarly if we suppress the propagation of $1^+$ by setting $h_{11}=-\tfrac{40}{17}h_7$, we remain with
\be
r_2=\frac{34}{75h_7}<0
\ ,\quad
r_-=-\frac{34(8B^2+25C^2)}{(16B+25C)^2h_7}<0
\ .
\ee
Again, these conditions cannot be satisfied simultaneously, but if we also remove the $1^-$ state by the condition
$C=-16B/25$, we remain just with the massless graviton and a massive $2^-$ state 
with squared mass $m_2^2=-\frac{17B}{75h_7}$.

In summary, there are just two solutions with the assumed properties, namely
\bea
\mathrm{Case\ I}\ :\quad&&
C=-\frac{16}{25}B\ ,\quad
h_{11}=-\frac{10}{3}h_7\ ,\quad
h_7<0\ ,\quad 
B>0\ ,\quad
\label{modelI}
\\
\mathrm{Case\ II}\ :\quad&&
C=-\frac{16}{25}B\ ,\quad
h_{11}=-\frac{40}{17}h_7\ ,\quad
h_7<0\ ,\quad 
B<0\ .
\label{modelII}
\eea
\smallskip
Finally, the saturated propagator is
\bea
\Pi&=&-\frac12\int d^4q\Biggl\{
{\cal J}
\left(\sum_{i,j=1,2,4} b_{ij}^{-1}(2^+)P_{ij}(2^+)
+\sum_{i,j=3,4,5} b_{ij}^{-1}(0^+)P_{ij}(0^+)\right){\cal J}
\\
&&\!\!\!\!\!\!\!
+\tau\bigg[ b^{-1}(3^-)\,P(3^-)
+b^{-1}(2^-)\,P_{11}(2^-)
+b^{-1}(1^+)\,P_{11}(1^+)
+\sum_{i,j=2,4,5}b^{-1}(1^-)_{ij}\,P_{ij}(1^-)\bigg]\,\tau
\Biggr\}\ .
\nn
\label{nf}
\eea
We can make this expression more understandable by explicitly displaying the denominators of each propagator, and evaluating the contractions
of the spin projectors with the sources:
\bea
\Pi&=&-\frac12\int d^4q\Biggl\{
-\frac{4}{A}\,\tau\cdot P(3^-,\eta)\cdot\tau
-\frac{1}{a_0 q^2} 
S_{ab}
\left(P_{44}^{abcd}(2^+,\eta)-\frac12 P_{55}^{abcd}(0^+,\eta) \right)
S_{cd}
\nn\\
&&
+\frac{16}{30h_7+9h_{11}}\,\frac{1}{q^2+m_2^2}\,\tau\cdot P_{11}(2^-,m_2^2)\cdot\tau
\nn\\
&&
-\frac{1}{2B}\,\frac{1}{q^2+m_+^2}\mathrm{div}_1\tau_{[ab]}
\left(\eta^{ac}+\frac{q^aq^c}{m_+^2}\right)
\left(\eta^{bd}+\frac{q^bq^d}{m_+^2}\right)
\mathrm{div}_1\tau_{[cd]}
\nn\\
&&
-\frac{1}{6B^2(16B+25C)^2(2h_7+h_{11})}\,\frac{1}{q^2+m_-^2}
Z_a\left(\eta^{ab}+\frac{q^aq^b}{m_-^2}\right)Z_b
\nn\\
&&
+\frac{1}{45}\left(\frac{1}{A}-\frac{5}{16B+25C}\right)\tr_{13}\tau_a\, \tr_{13}\tau^a
\Biggr\}\ ,
\label{tff}
\eea
where we defined
\bea
S_{ab}&=&2i\,\mathrm{div}_1\tau_{ab}-i\,\mathrm{div}_2\tau_{ab}-2\sigma_{ab}\ ,
\\
Z_a&=&(16B+25C)(2h_7+h_{11})
(\mathrm{div}_{12}\tau_a-\mathrm{div}_{13}\tau_a)
-2B(4B+5C)\tr_{13}\tau_a\ ,
\eea
and
\be
\mathrm{div}_1\tau_{ab}=q^c\tau_{cab}\ ,\quad 
\mathrm{div}_2\tau_{ab}=q^c\tau_{acb}\ ,
\nn
\ee
\be
\mathrm{div}_{12}\tau_a=q^bq^c\tau_{cba}\ ,\quad 
\mathrm{div}_{13}\tau_a=q^bq^c\tau_{cac}\ ,\quad 
\mathrm{tr}_{13}\tau_a=\tau_{ca}{}^c\ .
\ee
This manifestly shows that in Cases I and II the saturated propagator contains the standard spin $2^+$ graviton, together with either a massive $1^+$ or massive $2^-$ propagators, contracted with suitable combinations of sources.
In particular the spin $1^+$ propagator, that may be less familiar, is the standard one of a massive two-form potential, described by the Lagrangian
\be
{\cal L} = -\frac{1}{12} H_{\mu\nu\rho} H^{\mu\nu\rho} - \frac12 m^2 B_{\mu\nu} B^{\mu\nu}\ ,
\ee
where $H_{\mu\nu\rho} = 3\partial_{[\mu} B_{\nu\rho]}$. We also note that, unlike the case of spin $2^+$, the spin $2^-$ propagator cannot be written solely in terms of second rank tensor sources, as it necessarily requires the presence of the 3rd rank sources.

\section{Conclusions}

In this paper we have set up the machinery that is necessary
to analyze the spectrum of a general MAG theory.
In particular, we have constructed the spin projectors for a general
three-index tensor, and we have used them to rewrite
the wave operator for the most general, 28-parameter MAG.
Not surprisingly, this case turns out to be too complicated
to determine its spectrum, but it is possible to
do so in special subclasses of theories.
We have considered here theories that have either vanishing non-metricity,
recovering previously known results,
and theories with vanishing torsion.
In the latter case the theory depends on 17 parameters;
imposing projective invariance reduces this to 10 parameters
and imposing that there be no propagating spin $3^-$ 
and no massive propagating spin $2^+$ fields,
further reduces this number to 6. 
The resulting action is given in \eq{faction}. 
Absence of ghosts and tachyons results in the inequalities \eqref{notac} and \eqref{nogh}
on these six parameters. 
We find that there are two solutions to these conditions, given in \eq{modelI}
and \eq{modelII}, describing either a massive spin $2^{-}$ or massive spin $1^{+}$ state, in addition to the standard graviton.
Even within the torsion-free subtheory, relaxing the conditions of section 6.3 will lead to a much more complicated system.

With hindsight, the absence of ghosts and tachyons in these models
is related the fact that, when converting to the $R$, $\phi$
variables in the manner of equation \eq{Rphi}, they do not contain any terms
quadratic in curvature. For the same reason, these models are also non-renormalizable.
This is entirely analogous to the situation also pointed out in \cite{Sezgin:1979zf} for the nine parameter metric quadratic theories with torsion. 

It is important to stress that the metric and torsion-free cases
are kinematically distinct from the original general MAG
and that the ghost- and tachyon-free models we have found
are {\it not} special cases of the general MAG,
but only of the kinematically restricted models
where either torsion or nonmetricity are set to zero.

Also of some interest would be the study of models with propagating spin $3^-$.
It is known that the free massless spin $3$ theory can be embedded 
in linearized MAG \cite{Baekler:2006vw}, however the underlying linearized gauge symmetry
does not extend to the full theory.
It would be interesting to explore whether MAG can
describe a massive spin 3 field coupled to gravity.
We hope to return to these questions in the future.

\goodbreak

{\it Note added}: After this work appeared on the {\tt arXiv},
we have been informed
that the spin projectors for the general theory have also been
worked out in \cite{schimidt}, and that they agree with ours.

\subsection*{Acknowledgments}

ES thanks Mitya Ponomarev for useful discussions. RP would like to thank the Mitchell Institute for Fundamental Physics and Astronomy
and ES thanks SISSA for hospitality. This work is supported in part by NSF grants PHY-1521099 and PHY-1803875
and by Mitchell Institute for Fundamental Physics and Astronomy.

 \pagebreak

\begin{appendix}

\section{Spin projectors}
\label{proj}

In the torsion-free case the spin projectors have
also been given in \cite{Alvarez:2018nxc}.

\subsection{ $P(J^-)$ projectors, $J=0,1,2,3$ }

Let us introduce the notation
\bea
\Pi_i :=  ( \Pi^{(ts)},\  \Pi^{(hs)},\Pi^{(ha)},\ \Pi^{(ta)},\ \Pi^{(s)},\ \Pi^{^a} )\ ,\qquad i=1,2,...,6\ ,
\eea
where we recall that $\Pi^{(ts)},...,\Pi^{^a}$ are defined in \eq{psa}, \eq{sa3a} and \eq{sa3b}.

The negative parity projectors are given by
\bea
P(3^-) &=& \Pi_1 \left( TTT\right)  \Pi_1 - P(1^-)_{11}\ ,
\label{sp4}
\w6
P(2^-)_{i-1,j-1}   &=& \Pi_i A_{ij} \Pi_j -  P(1^-)_{ij}\ , \qquad i,j =2,3\ ,
\nn\w6
P(1^-)_{ij} &=& \Pi_i\,B_{ij}\,\Pi_j\ ,\qquad i,j=1,2,3\ ,
\nn\w4
P(1^-)_{i,3+j} &=& \Pi_i\,B_{i,3+j}\,\Pi_j\ ,\qquad i,j=1,2,3\ ,
\nn\w4
P(1^-)_{3+i,3+j} &=& \Pi_i\,B_{ij}\,\Pi_j\ ,\qquad i,j=1,2,3\ ,
\nn\w4
P(1^-)_{i7} &=& \Pi_i\,B_{i7}\,\Pi_5\ ,\qquad P(1^-)_{3+i,7} = \Pi_i\,B_{3+i,7}\,\Pi_5\ ,\quad i=1,2,3\ ,
\nn\w4
P(1^-)_{77} &=& \Pi_5 \left(TL+LT\right) \Pi_5\ ,
\nn\w6
P(0^-) &=& \Pi_4 \left(TTT\right) \Pi_4\ ,
\eea
where it is understood that there is no summation over the indices displayed, and 
\bea
A(2^-)_{ij} &:=& \begin{pmatrix} 1 & \sqrt 2 \\ \sqrt 2 & 1\\ \end{pmatrix} TTT\ ,
\w4
B(1^-)_{ij} &:=& \begin{pmatrix} 
\dfrac{3}{d+1} & \dfrac{3\sqrt 2}{\sqrt{(d-2)(d+1)}} & \dfrac{\sqrt 6}{\sqrt{(d-2)(d+1)}}\\
\dfrac{3\sqrt 2}{\sqrt{(d-2)(d+1)}} & \dfrac{6}{d-2} & \dfrac{2\sqrt 3}{d-2}\\
\dfrac{\sqrt 6}{\sqrt{(d-2)(d+1)}} & \dfrac{2\sqrt 3}{d-2} & \dfrac{2}{d-2}\\
\end{pmatrix} T_{12} T^{12} T\ ,
\nn
\eea
\bea
B(1^-)_{i,3+j} &:=& \begin{pmatrix} 
\dfrac{3}{\sqrt{d+1}} & \dfrac{3\sqrt 2}{\sqrt{d+1}} & \dfrac{\sqrt 6}{\sqrt{d+1}}\\
\dfrac{3\sqrt 2}{\sqrt{d+1}} & \dfrac{6}{\sqrt{d-2}} & \dfrac{2\sqrt 3}{\sqrt{d-2}}\\
  \dfrac{\sqrt 6}{\sqrt{d+1}} & \dfrac{2\sqrt 3}{\sqrt{d-2}} & \dfrac{2}{\sqrt{d-2}}\\
\end{pmatrix} L_{12}T^{12}T \ , 
\w4
B(1^-)_{3+i,3+j} &:=& 
\begin{pmatrix} 
LLT+LTL+TLL & 3{\sqrt 2}\,LLT & {\sqrt 6}\,LLT \\
3{\sqrt 2}\,LLT & LLT+LTL+TLL  & 2{\sqrt 3}\,LLT\\
{\sqrt 6}\,LLT & 2{\sqrt 3}\,LLT & LLT+LTL+TLL \\
\end{pmatrix}\ , 
\nn\w4
\left[B(1^-)_{i7}\right]_{cab}{}^{ef}  &:=&  \begin{pmatrix} 
\dfrac{\sqrt 6}{\sqrt {d+1}}\, T_{23}\, \hat q^{\prime}\, T \\ -\dfrac{\sqrt 3}{\sqrt{d-2}} \,T_{23}\, \hat q^{\prime}\, T  \\ \  \dfrac{2}{\sqrt{d-2}}\, T_{31}\, \hat q^{\prime}\, T \end{pmatrix}\ ,
\qquad 
\left[B(1^-)_{3+i,7}\right]_{cab}{}^{ef} := \begin{pmatrix} 
{\sqrt 6}\, L_{23}\, \hat q^{\prime}\, T  \\ -{\sqrt 3}\,L_{23}\, \hat q^{\prime}\, T  \\ \quad\ 2\, L_{31}\, \hat q^{\prime}\, T  \end{pmatrix}\ .
\nn
\eea
Note that the transposition raises and lowers the vector indices on $T$ and $L$ such that, for example, $T_c{}^d$ and $T_{ca}$ get mapped to $T_d{}^c$ and $T^{ca}$, respectively. Therefore, we have $(B^T)_{def}{}^{cab} = T_{de} L^{ca} T_f{}^b$.

\subsection{ $P(J^+)$ projectors, $J=0,1,2$ }

The positive parity projectors are given by
\bea
P(2^+)_{ij} &=& \Pi_i C_{ij} \Pi_j- P(0^+)_{ij}\ ,\quad P(2^+)= \Pi C_{i4} \Pi_5\ ,\quad P(2^+)_{44} = \Pi_5 \left( TT\right) \Pi_5 - P(0^+)_{55}\ ,
\nn\w4
P(1^+)_{i-1,j-1} &=& \Pi_i D_{ij} \Pi_j\ ,\qquad i,j=2,3,4\ ,
\nn\w4
P(0^+)_{ij} &=& \Pi_i\,E_{ij}\,\Pi_j\ ,\quad P(0^+)_{i4}= \Pi_i\, E_{i4}\,\Pi_1\ , \quad P(0^+)_{44} = \Pi_1 \left(LLL\right) \Pi_1\ ,
\nn\w4
P(0^+)_{ir} &=& \Pi_i\,E_{ir}\,\Pi_5\ ,\quad P(0^+)_{4r} = \Pi_1\, E_{4r}\,\Pi_5\ , \quad P(0^+)_{rs} = \Pi_5\, E_{rs}\,\Pi_5\ , 
\eea
\bea
C(2^+)_{ij}  &:=& \begin{pmatrix} 
TTL+TLT+LTT &&\dfrac{3}{2\sqrt 2}\, LTT && {\sqrt 6}\,TLT\\ 
\dfrac{3}{2\sqrt 2}\, LTT && \dfrac32\, LTT &&\Pi_5(TT)\,L \\ 
{\sqrt 6}\,TLT && \Pi_5(TT)\,L && TTL+TLT-\dfrac12 LTT \end{pmatrix}\ ,
\nn\w4
C(2^+)_{i4}  &:=& \begin{pmatrix}
{\sqrt 3}\,\hat q_c\,T_a{}^e\,T_b{}^f \\ \sqrt{\dfrac32}\,T_a{}^e\,T_b{}^f\\ \dfrac{1}{\sqrt 2}\,T_c{}^e\,\hat q_a\,T_b{}^f \end{pmatrix}\ ,
\nn\w4
D(1^+)_{ij}   &:= & \begin{pmatrix} 
TTL+TLT-\frac12 LTT && 2{\sqrt 3}\,\Pi_6(TT)\,L && {\sqrt 6}\,TTL\\ 
2{\sqrt 3}\,\Pi_6(TT)\,L && \dfrac32\, LTT && -\dfrac{3}{\sqrt 2}\,LTT \\ 
{\sqrt 6}\, TTL && -\dfrac{3}{\sqrt 2}\,LTT  && TTL+TLT+LTT \end{pmatrix}\ ,
\nn\w4
E(0^+)_{ij} &:=& \dfrac{1}{d-1} \begin{pmatrix} 
3\,LT_{23}T^{23}  & \dfrac{3}{\sqrt 2}\,LT_{23}T^{23} & {\sqrt 6}\,T_{31}T_{31}L\\
\dfrac{3}{\sqrt 2}\,LT_{23}T^{23} & \dfrac32\,LT_{23}T^{23} & 2{\sqrt 3}\, T_{12}T^{12}L\ , \\
{\sqrt 6}\,T_{31}T_{31}L & 2{\sqrt 3}\, T_{12}T^{12}L & 2\, T_{12}T^{12}L
\end{pmatrix}\ , 
\nn\w4
E(0^+)_{i4} &:=& \dfrac{1}{\sqrt {d-1}} \begin{pmatrix}
{\sqrt 3}\,LT_{23}T^{23}  \\ \sqrt{\dfrac 32}\,LT_{23}T^{23}  \\ -{\sqrt 2}\, T_{12}T^{12}L 
\end{pmatrix}\ ,
\nn
\eea
\bea
E(0^+)_{ir} &:= & \begin{pmatrix}
\dfrac{\sqrt 3}{d-1}\,\hat q\,T_{23}\,T^{23} & \dfrac{\sqrt 3}{\sqrt{d-1}}\,\hat q\,T_{23}\,L^{23}\\
\dfrac{\sqrt 3}{2(d-1)}\,\hat q\,T_{23}\,T^{23} & \dfrac{\sqrt 3}{\sqrt{d-1}}\,\hat q\,T_{23}\,L^{23}\\ 
\dfrac{\sqrt 3}{d-1}\,\hat q\,T_{31}\,T^{12} & \dfrac{\sqrt 2}{\sqrt{d-1}}\,\hat q\,T_{31}\,L^{12}
\end{pmatrix}\ ,
\qquad
E(0^+)_{4r}   := \begin{pmatrix} 
\dfrac{\sqrt 2}{\sqrt{d-1}}\, L_{12}\,\hat q\,T^{23}\\ \qquad\quad\  L_{12}\,\hat q\,L^{23}
\end{pmatrix}\ ,
\nn\w4
E(0^+)_{rs} &:=& \begin{pmatrix} 
\dfrac{1}{d-1}\,T_{23}\,T^{23} & \dfrac{1}{\sqrt{d-1}}\,T_{23}\,L^{23}\\ \dfrac{1}{\sqrt{d-1}}\,L_{23}\,T^{23} & \qquad\quad \  L_{23}\,L^{23} 
\end{pmatrix}\ ,\qquad i=1,2,3, \quad r,s=5,6\ .
\eea

 \newpage

\section{The coefficient matrices}
\label{cm1}

\subsection{General MAG}

Here we provide the coefficient matrices $a(J^P)$ arising in the expansion of the wave operator in the general 28 parameter model, in terms of the spin projection operators.  
As a weak check, we observe that all coefficient matrices vanish identically
for the combination \eqref{gaussbonnet}.
\bea
a(3^-) &=& (2c_1+2c_2+c_4+c_5+c_6)(-q^2)-a_0-4a_4-4a_5
\w4
a(2^-)_{11}&=&  2 (c_1+c_2-c_4-c_5-c_6) (-q^2)
+\frac12 \left( a_0-6a_1- 3a_2 - 8a_4+4a_5 -6a_9\right)
\nn\\
a(2^-)_{12} &=&
\frac{\sqrt 3}{2}\left[(c_4-c_6)(-q^2)-2a_1-a_2-a_9\right]
\nn\\
a(2^-)_{22}&=&
\left(2(c_1-c_2)+\frac12(c_4-c_5+c_6)\right) (-q^2)+\frac12 a_0-a_1-\frac12 a_2
\eea
\bea
a(2^+)_{11}&=&\left(\frac23(2c_1+2c_2+c_4+c_5+c_6)+\frac13(c_7+c_8+c_9+c_{10}+c_{11}+c_{12})\right)(-q^2)
\nn\\
&& -a_0-4a_4-4a_5
\nn\\
a(2^+)_{12} &=&-\frac{1}{3\sqrt 2}\big[2 (2 c_1 + 2 c_2 + c_4 + c_5 + c_6)
 +c_7 + c_8 + c_9 + c_{10} + c_{11} + c_{12}\big](-q^2)
\nn\\
a(2^+)_{13} &=&\frac{1}{\sqrt 6}
(c_7 + c_8 - c_9 - c_{10})(-q^2)
\nn\\
a(2^+)_{14} &=&\frac{1}{2\sqrt3}
(a_0+4a_4+4a_5)i|q|
\nn\\
a(2^+)_{22}&=&\left(\frac13(2c_1+2c_2+c_4+c_5+c_6)
+\frac16(c_7+c_8+c_9+c_{10}+c_{11}+c_{12})\right) (-q^2)
\nn\\
&& +\frac12 a_0-3a_1-\frac32a_2-4a_4+2a_5-3a_9
\nn\\
a(2^+)_{23} &=&\frac{1}{2\sqrt 3}
\left[(-c_7 - c_8 + c_9 + c_{10})(-q^2)
-6a_1-3a_2-3a_9)\right]
\nn\\
a(2^+)_{24} &=&\frac{1}{2\sqrt6}
(-a_0+8 a_4 - 4 a_5 + 3 a_9)i|q|
\nn\\
a(2^+)_{33}&=&\left((2c_1-2c_2+2c_3+c_4-c_5+c_6)
+\frac12(c_7+c_8+c_9+c_{10}-c_{11}-c_{12})\right)(-q^2)
\nn\\
&& +\frac12 a_0-a_1-\frac12 a_2
\nn\\
a(2^+)_{34} &=&\frac{1}{2\sqrt2}(a_0+a_9)i|q|
\nn\\
a(2^+)_{44}&=&-a_4 q^2
\eea

\bea
a(1^-)_{11} &=&
\left(2 c_1 + 2 c_2 + c_4 + c_5 + c_6
+\frac53(c_7+c_9+c_{11}+2c_{13}-c_{14}-c_{15})\right)(-q^2)
\nn\\
&&
+\frac23 a_0 -4a_4-4a_5
-\frac{20}{3}(a_6+a_7+a_8)
\nn\\
a(1^-)_{12} &=&\frac{\sqrt5}{6}
\Big[\left(2 c_7 + 2 c_9 + 2 c_{11} - 8 c_{13} + c_{14} + c_{15}\right)(-q^2)
+2a_0+16 a_6-8a_7+4a_8-6a_{10}-6a_{11}\Big]
\nn\\
a(1^-)_{13} &=&\frac{\sqrt 5}{2\sqrt3}
\Big[\left(2 c_7 - 2 c_9 - c_{14} + c_{15}\right)(-q^2)
-2a_{10}-2a_{11}\Big]
\nn\\
a(1^-)_{14} &=&\frac{\sqrt{5}}{3}
\Big[(-c_8-c_{10}-c_{12}+2 c_{13}-c_{14}- c_{15})(-q^2)
+a_0-4a_6-4 a_7-4a_8\Big]
\nn\\
a(1^-)_{15} &=&\frac{\sqrt{5}}{\sqrt{18}}
\Big[2 (c_8 + c_{10} + c_{12} - 2 c_{13} + c_{14} + c_{15})(-q^2)
+a_0+8 a_6-4 a_7+2 a_8-3 a_{10}-3 a_{11}\Big]
\nn\\
a(1^-)_{16} &=&-\frac{\sqrt{5}}{\sqrt{6}}
(a_{10} + a_{11})
\nn\\
a(1^-)_{17} &=&\sqrt{\frac{5}{24}}(-a_0+4a_7+2a_8)i|q|
\nn\\
a(1^-)_{22} &=&\frac{1}{6}
(12 c_1 + 12 c_2 -3 c_4 - 3 c_5 - 3 c_6
+2 c_7 + 2 c_9 + 2 c_{11} + 16 c_{13} + 4 c_{14} + 4 c_{15}) (-q^2)
\nn\\
&&
+\frac56 a_0-3a_1-\frac32a_2 
-3a_3-4a_4+2a_5-\frac{16}{3} a_6-\frac43 a_7
+\frac83a_8-3a_9+4a_{10}-2a_{11} 
\nn\\
a(1^-)_{23} &=&\frac{1}{2\sqrt3}
\left[(3c_4-3c_6+2 c_7 - 2 c_9 + 2 c_{14} - 2 c_{15})(-q^2)
-6a_1-3a_2-6a_3-3a_9+4 a_{10}-2 a_{11}\right]
\nn\\
a(1^-)_{24} &=&\frac16
\left[(-2 c_8 - 2 c_{10} - 2 c_{12} - 8 c_{13} + c_{14} + c_{15}) (-q^2)
+2a_0
+16 a_6-8a_7+4 a_8-6a_{10}-6a_{11}\right]
\nn\\
a(1^-)_{25} &=&\frac{1}{3\sqrt2}
\big[(2 c_8 + 2 c_{10} + 2 c_{12} + 8 c_{13} - c_{14} - c_{15})(- q^2)
\nn\\
&& +a_0-9 a_3-16 a_6-4 a_7+8 a_8+12 a_{10}-6 a_{11}
\big]
\nn\\
a(1^-)_{26} &=&\frac{1}{\sqrt6}(-3a_3+2 a_{10}- a_{11})
\nn\\
a(1^-)_{27} &=&\frac{1}{2\sqrt6}
(-a_0+4 a_7-4 a_8+3 a_{11})i|q|
\nn\\
a(1^-)_{33} &=&\frac{1}{2}
\left[(4 c_1 - 4 c_2 + c_4 - c_5 + c_6+2 c_7 + 2 c_9 - 2 c_{11})(-q^2)
-a_0-2 a_1-a_2-2 a_3\right]
\nn\\
a(1^-)_{34} &=&\frac{1}{2\sqrt3}
\left[(-2 c_8 + 2 c_{10} - c_{14} + c_{15}) (-q^2)-2a_{10}-2a_{11}
\right]
\nn\\
a(1^-)_{35} &=&\frac{1}{\sqrt6}
\left[(2 c_8 - 2 c_{10} + c_{14} - c_{15})(-q^2)
-3a_3+2 a_{10}-a_{11}\right]
\nn\\
a(1^-)_{36} &=&-\frac{1}{\sqrt2}(a_0+a_3)
\nn\\
a(1^-)_{37} &=&\frac{1}{2\sqrt2}(-a_0+a_{11})i|q|
\nn
\eea

\bea
a(1^-)_{44} &=&\frac{1}{3}
\Big[(2c_1+2c_2+c_4+c_5+c_6+c_7+c_9+c_{11}+2c_{13}-c_{14}-c_{15}) (-q^2)
\nn\\&&
-2a_0-12a_4-12 a_5-4a_6-4a_7-4a_8
\Big]
\nn\\
a(1^-)_{45} &=&\frac{1}{3\sqrt2}
\Big[-2(2c_1+2c_2+c_4+c_5+c_6+c_7+c_9+c_{11}+2c_{13}-c_{14}-c_{15})(-q^2)
\nn\\&&
+a_0+8 a_6- 4 a_7+ 2 a_8-3 a_{10}-3 a_{11}
\Big]
\nn\\
a(1^-)_{46} &=&-\frac{1}{\sqrt6}(a_{10}+a_{11})
\nn\\
a(1^-)_{47} &=&\frac{1}{2\sqrt6}(a_0+8 a_4 +8 a_5 +4 a_7 + 2a_8) i|q|
\nn\\
a(1^-)_{55} &=&\frac{2}{3}
(2c_1+2c_2+c_4+c_5+c_6+c_7+c_9+c_{11}+2c_{13}-c_{14}-c_{15})(-q^2)
\nn\\&&
+\frac23 a_0
-3a_1-\frac32 a_2-\frac32 a_3-4a_4 +2a_5-\frac83a_6-\frac23 a_7 
+\frac43a_8-3a_9+2a_{10}- a_{11}
\nn\\
a(1^-)_{56} &=&\frac{1}{2\sqrt3}
(-6a_1-3a_2-3a_3-3a_9+2a_{10}-a_{11})
\nn\\
a(1^-)_{57} &=&\frac{1}{4\sqrt3}
\left[-2a_0+8a_4-4a_5+4 a_7-4 a_8+3a_9+3a_{11}\right]i|q|
\nn\\
a(1^-)_{66} &=&-\frac{1}{2}(2a_1+a_2+a_3)
\nn\\
a(1^-)_{67} &=&\frac14(a_9+a_{11})i|q|
\nn\\
a(1^-)_{77} &=&\frac12(2a_4+a_5+a_7)(-q^2)
\eea

\bea
a(1^+)_{11}&=&\frac12\big[(4c_1+4c_2+c_7-c_8+c_9-c_{10}+c_{11}-c_{12})(-q^2)
\nn\\
&&+a_0-6 a_1-3 a_2-8 a_4+4 a_5-6 a_9\big]
\nn\\
a(1^+)_{12} &=&-\frac{1}{2\sqrt3}
\big[
\left(2c_4-2c_6+c_7-c_8-c_9+c_{10}\right)(-q^2)
-6a_1-3a_2-3a_9\big]
\nn\\
a(1^+)_{13} &=&\frac{1}{\sqrt 6}
(-2c_4+2c_6-c_7+c_8+c_9-c_{10})(-q^2)
\nn\\
a(1^+)_{22}&=&\frac16\left(4c_1-4c_2-4c_3+
c_7 - c_8 + c_9 - c_{10} - c_{11} + c_{12}\right)(-q^2)
+\frac12 a_0-a_1-\frac12 a_2
\nn\\
a(1^+)_{23} &=&\frac{1}{3\sqrt2}
(4 c_1 - 4 c_2 - 4 c_3
+c_7 - c_8 + c_9 - c_{10} - c_{11} + c_{12})(-q^2)
\nn\\
a(1^+)_{33}&=&\frac13\left(4 c_1 - 4 c_2 - 4 c_3
+c_7 - c_8 + c_9 - c_{10} - c_{11} + c_{12}\right)(-q^2)
\nn\\
&& -a_0-4a_1+4a_2
\eea

\bea
a(0^+)_{11}&=&\frac23(2 c_1 + 2 c_2 + c_4 + c_5 + c_6
+2 c_7 + 2 c_8 + 2 c_9 + 2 c_{10} + 2 c_{11} + 2 c_{12})(-q^2)
\nn\\
&&-4 (a_4 + a_5 + a_6 + a_7 + a_8)
\nn\\
a(0^+)_{12} &=&\frac{1}{3\sqrt2}
\big[-2(2 c_1 + 2 c_2 + c_4 + c_5 + c_6
+2 c_7 + 2 c_8 + 2 c_9 + 2 c_{10} + 2 c_{11} + 2 c_{12})(-q^2)
\nn\\
&&-3a_0-24a_6+12a_7-6a_8+9 a_{10}+9 a_{11}
\big]
\nn\\
a(0^+)_{13} &=&\frac{1}{\sqrt6}
\left[2(-c_7 - c_8 + c_9 + c_{10}) (-q^2)
+3 (a_{10} + a_{11})\right]
\nn\\
a(0^+)_{14} &=&a_0-4(a_6 + a_7 + a_8)
\nn\\
a(0^+)_{15} &=&\frac{1}{2\sqrt3}
(a_0+4a_4 +4a_5 +12 a_6 +6 a_8) i|q|
\nn\\
a(0^+)_{16} &=&
\left(-\frac12a_0+2a_6 +2a_7 +2a_8\right) i|q|
\nn\\
a(0^+)_{22}&=&\frac13(2 c_1 + 2 c_2 + c_4 +
c_5+c_6+2 c_7 + 2 c_8 + 2 c_9 + 2 c_{10} + 2 c_{11} + 2 c_{12})(-q^2) 
\nn\\
&&+a_0-3 a_1-\frac32a_2-\frac92 a_3- 4 a_4+ 2 a_5
- 8 a_6- 2 a_7+4 a_8-3 a_9+6 a_{10}-3 a_{11}
\nn\\
a(0^+)_{23} &=&\frac{1}{2\sqrt3}
\left[
2(c_7+c_8-c_9-c_{10}) (-q^2)
-6a_1-3(a_2+3a_3+a_9)+6a_{10}-3a_{11}\right]
\nn\\
a(0^+)_{24} &=&\frac{1}{\sqrt2}\left[-a_0
-8 a_6+ 4 a_7-2 a_8+3 a_{10}+ 3 a_{11}\right]
\nn\\
a(0^+)_{25} &=&\frac{1}{2\sqrt6}
(-a_0+8a_4 - 4 a_5 +24 a_6 - 6 a_8+3 a_9-9 a_{10})i|q|
\nn\\
a(0^+)_{26} &=&\frac{1}{2\sqrt2}
(a_0+8 a_6-4 a_7+2 a_8-3 a_{10}-3 a_{11})i|q|
\nn\\
a(0^+)_{33}&=&(2 c_1 - 2 c_2 + 2 c_3 + c_4 - c_5 + c_6
+2 c_7 + 2 c_8 + 2 c_9 + 2 c_{10} - 2 c_{11} - 2 c_{12}+ 6 c_{16} ) (-q^2)
\nn\\
&&
-a_0-a_1-\frac12 a_2-\frac32 a_3
\nn\\
a(0^+)_{34} &=&\sqrt{\frac32}
(a_{10} + a_{11})
\qquad\qquad
a(0^+)_{35} =\frac{1}{2\sqrt2}
(-2a_0+a_9-3 a_{10})i|q|
\nn\\
a(0^+)_{36} &=&-\frac{\sqrt3}{2\sqrt2}
(a_{10} + a_{11})i|q|
\nn\\
a(0^+)_{44}&=&-4(a_4 + a_5 + a_6 + a_7 + a_8)
\nn\\
a(0^+)_{45} &=&\sqrt3 (2 a_6 + a_8) i|q|
\qquad\qquad
a(0^+)_{46} =
2(a_4 + a_5 + a_6 + a_7 + a_8)i|q|
\nn\\
a(0^+)_{55}&=&(a_4 + 3 a_6) (-q^2)
\qquad\qquad
a(0^+)_{56} =\frac{\sqrt3}{2}
(2a_6+a_8)(-q^2)
\nn\\
a(0^+)_{66}&=&(a_4 + a_5 + a_6 + a_7 + a_8) (-q^2)\,
\w4
a(0^-) &=& (2c_1-2c_2-c_4+c_5-c_6)(-q^2)
-a_0-4a_1+4a_2
\eea

\newpage

\subsection{Metric MAG}

Using \eq{metricc} and \eq{metrica} in the coefficient matrices of Appendix B.1,
and deleting the rows and columns that pertain to symmetric three-tensors,
we recover the coefficient matrices of the metric theory,
as computed in \cite{Sezgin:1979zf} (see also \cite{Floreanini:1993na}) with two differences.
First, one has to keep in mind that the graviton field $h_{ab}$ used here is equal to one half of the graviton field $\varphi_{ab}$ used in those references.
This gives a factor 2 in the mixed $A$-$h$ coefficients and 4 in the $h$-$h$ coefficients.
Second, the projectors $P_{ij}(1^+)$ with $i,j=2,3$
span the same space as $P_{ij}(1^+)$ with $i,j=1,2$ in those references, but differ by a linear transformation
(The old projectors do not respect the $GL(4)$ decomposition). This is of no consequence for the physical results.
\bea
a(2^-) &=&\frac12\left(\left(4g_1+g_4\right)(-q^2)
+a_0-2b_1-b_2\right)
\w4
a(2^+)_{33}&=&\frac12\left((4 g_1 + 4 g_3 + 2 g_4 + g_7 + g_8)(-q^2)+a_0 - 2 b_1 - b_2\right)
\nn\\
a(2^+)_{34} &=&\frac{i|q|}{2\sqrt2}(a0 + 2 b_1 + b_2)
\ ,\quad
a(2^+)_{44}=\frac{1}{4}(2b_1+b_2)(-q^2)
\w4
a(1^+)_{22}&=&\frac16 (4 g_1 - 4 g_3 + g_7 - g_8)(-q^2)
+\frac12 a_0-2b_1+b_2
\nn\\
a(1^+)_{23} &=&\frac{4 g_1 - 4 g_3 + g_7 - g_8}{3\sqrt2}(-q^2)
\nn\\
a(1^+)_{33} &=&\frac13 (4 g_1 - 4 g_3 + g_7 - g_8)(-q^2)-a_0 + 8 b_1 - 8 b_2
\w4
a(1^-)_{33}&=&\frac12\left((4 g_1 + g_4 + 2 g_7)(-q^2)
-a_0 - 2 b_1 - b_2 - 2 b_3\right)
\nn\\
a(1^-)_{36} &=&-\frac{a_0 + b_3}{\sqrt2}
\ ,\qquad\qquad\ \
a(1^-)_{37} =\frac{i|q|}{2\sqrt2}(-a_0 + b_3)
\nn\\
a(1^-)_{66} &=&-\frac12(2 b_1+b_2+b_3)
\ ,\quad
a(1^-)_{67} =\frac{i|q|}{4} (2 b_1 + b_2 + b_3)
\nn\\
a(1^-)_{77} &=&\frac18(2 b_1 + b_2 + b_3)(-q^2)
\w4
a(0^+)_{33}&=&(2 g_1 + 2 g_3 + g_4 + 2 g_7 + 2 g_8 + 6 g_{16})(-q^2)
+\frac12 (-2 a_0 - 2 b_1 - b_2 - 3 b_3)
\nn\\
a(0^+)_{35} &=&-\frac{i|q|}{2\sqrt2}(2 a_0 - 2 b_1 - b_2 - 3 b_3)
\nn\\
a(0^+)_{55}&=&\frac{1}{4}(2 b_1 + b_2 + 3 b_3)(-q^2) 
\nn\\
a(0^+)_{36}&=&a(0^+)_{56}=a(0^+)_{66}=0
\w4
a(0^-) &=& (2g_1-g_4)(-q^2) -a_0 - 4 b_1 + 4 b_2
\eea

\newpage

\subsection{Torsion-free MAG}

We give here the coefficient matrices for the
most general torsion-free model, as discussed in section 6.1.
If one wishes to further impose the projective symmetry
discussed in Sect. 6.2, one has to further impose (in four dimensions)
the conditions given in equation (\ref{tfp}).

{\it Beware of our notational convention:} the indices $i,j$ on the coefficient matrix 
$a_{ij}(J^\cP)$ refer to the representations they carry.
Thus, they do not always agree with the usual convention
of numbering matrix elements.
For example, the representation $2^+_3$ is absent from the
symmetric tensor $A_{cab}$; only the representations $1,2,4$
are present. Accordingly
the element of $a(2^+)$ in the third row and column
is labelled $a_{44}(2^+)$.

\bea
a(3^-)&=&(2h_1+2h_2+h_4)(-q^2)-a_0-4a_4-4a_5
\label{sp3tfp}
\\
a(2^+)_{11}&=&\frac13\left(4h_1+ 4 h_2 + 2 h_4 + h_7 + h_8 + h_9+h_{10} + h_{11} + h_{12}\right)(-q^2)
\nn\\
&& -a_0-4a_4-4a_5
\nn\\
a(2^+)_{12} &=&\frac{1}{6\sqrt 2}\big[4 h_1 + 4 h_2 + 2 h_4 + 4 h_7 + 4 h_8 - 2 h_9
- 2 h_{10} + h_{11} + h_{12} \big](-q^2)
\nn\\
a(2^+)_{14} &=&\frac{1}{2\sqrt3}
(a_0+4a_4+4a_5)i|q|
\nn\\
a(2^+)_{22}&=&\frac{1}{6}\left(10 h_1  - 8 h_2 + 9 h_3 + 5 h_4 + 4 h_7 + 4h_8 + h_9+ h_{10} - 2 h_{11} - 2 h_{12}\right) (-q^2)
\nn\\
&& +\frac12 (a_0-2a_4+a_5)
\nn\\
a(2^+)_{24} &=&\frac{1}{\sqrt6}
(a_0-2a_4+a_5)i|q|\ ,\qquad
a(2^+)_{44}=a_4 (-q^2)\ .
\\
a(2^-)_{11}&=&\left(2 h_1-h_2-\frac12 h_4\right)(-q^2)
+\frac12 a_0-a_4+\frac12 a_5
\\
a(1^+)_{11}&=&\frac12\big[(2h_1-h_3-h_4+h_9-h_{10})
(-q^2)
+a_0-2a_4+a_5\big]
\eea

\bea
a(1^-)_{11} &=&
\frac13\left(6h_1 +6h_2 +3h_4 +5h_7+5h_9+5h_{11}\right)(-q^2)
+\frac23 a_0 -4a_4-4a_5
-\frac{20}{3}(a_6+a_7+a_8)
\nn\\
a(1^-)_{12} &=&-\frac{\sqrt5}{6}
\Big[\left(-2h_7 + 4h_9 +h_{11}\right)(-q^2)
+a_0+8 a_6-4a_7+2a_8\Big]
\nn\\
a(1^-)_{14} &=&\frac{\sqrt{5}}{3}
\Big[(-h_8-h_{10}-h_{12})(-q^2)
+a_0-4a_6-4 a_7-4a_8\Big]
\nn\\
a(1^-)_{15} &=&-\frac{\sqrt{5}}{6\sqrt{2}}
\Big[2(h_8+h_{10}+h_{12})(-q^2)
+a_0+8 a_6-4 a_7+2 a_8\Big]
\nn\\
a(1^-)_{17} &=&{\frac{\sqrt5}{2\sqrt6}}(-a_0+4a_7+2a_8)i|q|
\eea
\bea
a(1^-)_{22} &=&\frac{1}{6}\left[
(12h_1-6h_2-3 c_4+2h_7-8h_9 +4h_{11})(-q^2)
-a_0-6a_4+3a_5-8a_6-2a_7+4a_8\right]
\nn\\
a(1^-)_{24} &=&-\frac{1}{6}
\left[(2h_8-4h_{10}-h_{12})(-q^2)
+a_0
+8a_6-4a_7+2a_8\right]
\nn\\
a(1^-)_{25} &=&\frac{1}{6\sqrt2}
\big[(-2h_8 +4h_{10} +h_{12})(- q^2)
-4a_0-8a_6-2a_7+4a_8\big]
\nn\\
a(1^-)_{27} &=&-{\frac{1}{2\sqrt6}}(a_0+2a_7-2a_8)i|q|
\nn\\
a(1^-)_{44} &=&\frac{1}{3}
\Big[(2h_1+2h_2+h_4+h_7+h_9+h_{11})(-q^2)
-2a_0-12a_4-12 a_5-4a_6-4a_7-4a_8
\Big]
\nn\\
a(1^-)_{45} &=&\frac{1}{6\sqrt2}
\Big[2(2h_1+2h_2+h_4+h_7+h_9+h_{11})(-q^2)
-a_0-8 a_6+4 a_7-2 a_8\Big]
\nn\\
a(1^-)_{47} &=&\frac{1}{2\sqrt6}
(a_0+8a_4+8a_5+4a_7+2a_8)i|q|
\nn\\
a(1^-)_{55} &=&\frac{1}{6}\left[
(2h_1+2h_2+h_4+h_7+h_9+h_{11})(-q^2)
+ a_0-6a_4 +3a_5-4a_6-a_7+2a_8\right]
\nn\\
a(1^-)_{57} &=&\frac{1}{4\sqrt3}
(a_0-4a_4+2a_5-2a_7+2a_8)i|q|
\nn\\
a(1^-)_{77} &=&\frac12(2a_4+a_5+a_7)(-q^2)
\eea

\bea
a(0^+)_{11}&=&\frac23(2 h_1 + 2 h_2 +h_4 
+ 2 h_7 + 2 h_8 + 2h_9 + 2h_{10} + 2h_{11} + 2h_{12})(-q^2)
\nn\\
&&-4 (a_4 + a_5 + a_6 + a_7 + a_8)
\nn\\
a(0^+)_{12} &=&\frac{1}{6\sqrt2}
\big[2(2h_1 + 2h_2 +h_4-h_7-h_8 +5h_9 +5h_{10} + 2 h_{11} + 2h_{12})(-q^2)
\nn\\
&&+3a_0+24a_6-12a_7+6a_8
\big]
\nn\\
a(0^+)_{14} &=&a_0-4(a_6 + a_7 + a_8)
\nn\\
a(0^+)_{15} &=&\frac{1}{2\sqrt3}
(a_0+4a_4+4a_5 +12a_6+6 a_8) i|q|
\ ,\qquad
a(0^+)_{16} =\frac12
(-a_0+4a_6+4a_7+4a_8)i|q|
\nn\\
a(0^+)_{22}&=&\frac{1}{6}(10h_1-8h_2+9h_3 + 5h_4 +7h_7 +7h_8 +13h_9 +13h_{10}-8h_{11}-8h_{12}+27h_{16})(-q^2) 
\nn\\
&&+\frac12\left(-a_0-2a_4+ a_5
-4a_6- a_7+2a_8\right)
\nn\\
a(0^+)_{24} &=&\frac{1}{2\sqrt2}
\left[a_0+8 a_6-4 a_7+2 a_8\right]
\ ,\qquad
a(0^+)_{25} =-\frac{1}{4\sqrt6}
(5a_0+8a_4 -4a_5 +24a_6-6 a_8) i|q|
\nn\\
a(0^+)_{26} &=&-\frac{1}{4\sqrt2}\left(a_0+8a_6-4a_7 +2a_8\right) i|q|
\nn\\
a(0^+)_{44}&=&-4(a_4 + a_5 + a_6 + a_7 + a_8)
\qquad\quad
a(0^+)_{45} =\sqrt3(2a_6+a_8)i|q|
\nn\\
a(0^+)_{46} &=&2(a_4+a_5+a_6+a_7+a_8)i|q|
\ ,\qquad
a(0^+)_{55}=(a_4 + 3 a_6) (-q^2)
\nn\\
a(0^+)_{56} &=&\frac{\sqrt3}{2}
(2a_6+a_8)(-q^2)
\ ,\qquad
a(0^+)_{66}=(a_4 + a_5 + a_6 + a_7 + a_8) (-q^2)\,
\w4
a(0^-) &=&  0
\eea

\end{appendix}

\end{document}